\documentclass[aps,reprint,twocolumn,floatfix,
groupedaddress]{revtex4-1}

\usepackage{graphicx}
\usepackage{dcolumn} 
\usepackage{bm}      
\usepackage{siunitx}
\usepackage{curves}
\usepackage{epic}

\usepackage{amsthm}
\usepackage{amsmath}
\usepackage{amssymb}

\usepackage{newtxtext}
\usepackage[smallerops]{newtxmath}

\usepackage{microtype}
\usepackage[export]{adjustbox}
\usepackage{physics}
\usepackage{hyperref}
\usepackage[caption=false]{subfig}
\usepackage{wasysym}
\usepackage{color}

\usepackage{tabularx}
\usepackage{booktabs}
\usepackage{multirow}

\usepackage{setspace}

\definecolor{darkerblue}{RGB}{33, 85, 168}

\hypersetup{
    colorlinks = true,
    linkcolor  = darkerblue,
    citecolor  = darkerblue,
    urlcolor   = darkerblue,
}
\setcitestyle{numbers,square}

\newcommand{\dtoprule}{\specialrule{1pt}{0pt}{0.4pt}%
            \specialrule{0.3pt}{0pt}{\belowrulesep}%
            }
\newcommand{\dbottomrule}{\specialrule{0.3pt}{0pt}{0.4pt}%
            \specialrule{1pt}{0pt}{\belowrulesep}%
            }

\def\openone{\leavevmode\hbox{\small1\kern-4.2pt\normalsize1}}

\def\DTO{Dy\textsubscript{2}Ti\textsubscript{2}O\textsubscript{7}}
\def\HTO{Ho\textsubscript{2}Ti\textsubscript{2}O\textsubscript{7}}
\def\v#1{\boldsymbol{#1}}

\newcommand{\beq}{\begin{equation}}
\newcommand{\eeq}{\end{equation}}
\newcommand{\bea}{\begin{eqnarray}}
\newcommand{\eea}{\end{eqnarray}}
\newcommand{\bfig}{\begin{figure}}
\newcommand{\efig}{\end{figure}}
\newcommand{\bei}{\begin{itemize}}
\newcommand{\eei}{\end{itemize}}

\newcommand{\mc}[1]{\mathcal{#1}}

\begin{document}

\title{Long-range Coulomb interactions and nonhydrodynamic behavior in thermal quenches in spin ice}

\author{
Oliver Hart
}

\author{
Marianne Haroche
}

\author{
Claudio Castelnovo
}
\affiliation{
T.C.M.~Group, Cavendish Laboratory, JJ Thomson Avenue, Cambridge CB3 0HE, United Kingdom
            }

\date{June 2019}

\begin{abstract}
\setstretch{1.1}
When spin ice systems undergo a sudden thermal quench, they have been shown to enter long-lived metastable states where the monopole excitations form so-called noncontractible pairs [\href{http://dx.doi.org/10.1103/PhysRevLett.104.107201}{Phys.~Rev.~Lett.~\textbf{104}, 107201 (2010)}].
While the nature of these states is well understood, the dynamical mechanisms underpinning their formation remain largely unexplored and are the subject of this study.
We find that the long-range tail of the Coulomb interactions between monopoles plays a central role
by suppressing the monopole-assisted decay of noncontractible pairs with respect to monopole--antimonopole annihilation.
In conjunction with low final quench temperatures, where the system enters a non-hydrodynamic regime in which the monopoles effectively move at terminal velocity in the direction of the local force acting on them, the interactions lead to a metastable plateau that persists in the thermodynamic limit.
This is a remarkable phenomenon, reminiscent of jamming and some instances of glassiness: A transient modification of the short-time dynamics of the system allows it to enter a metastable state
whose lifetime can easily be astronomically large at (experimentally relevant) low temperatures.
We demonstrate this using Monte Carlo simulations and mean field population dynamics theory, and we provide an analytical understanding of the mechanisms at play.
When the interactions between monopoles are truncated to finite range, the metastable plateau reduces to a finite size effect.
We derive the finite size scaling behaviour of the density of noncontractible pairs in the metastable plateau for both short- and long-range interactions, and discuss the experimental implications of our results.
\end{abstract}
\maketitle
%
%

\section{\label{sec: intro}
Introduction
        }
Spin ice materials~\cite{Bramwell2001} are a class of three-dimensional frustrated magnets endowed at low temperature with topological properties and an emergent gauge symmetry~\cite{Castelnovo2012}. Moreover, they harbour collective excitations that take the form of itinerant, pointlike defects carrying a net magnetic charge: magnetic monopoles~\cite{Castelnovo2008}.
The nonequilibrium behaviour of these systems is particularly rich and exciting
and they can exhibit remarkably long relaxation and response timescales at low temperatures. While a number of attempts have been made to model and understand the origin of the dynamical behaviour in spin ice materials,
the complete picture arguably remains beyond our grasp.

In this paper, we make progress by investigating the specific setting of thermal quenches in classical spin ice~\cite{Castelnovo2010}, where these systems have been shown to enter long-lived metastable states in which the monopole excitations form so-called noncontractible pairs~\footnote{A noncontractible pair corresponds to a pair of oppositely charged monopoles residing on adjacent tetrahedra which are unable to annihilate by flipping the intervening spin.} (see Fig.~\ref{fig: Coulomb vs dipolar}).
While the nature of these states is well understood, the dynamical mechanisms underpinning their formation remain hitherto unexplored and are the subject of this work.
Using a combination of numerical simulations
and analytical mean field theory, we are able to provide a complete understanding of the phenomenon.
We find that the emergence of the plateau is rooted in two key ingredients: (i) the long-range nature of the Coulomb interaction between the monopoles and (ii) the fact that low temperature thermal quenches in spin ice can give rise to a non-hydrodynamic regime that increases the decay rate of the free monopole density in the system.
The latter feature is notably reminiscent of jamming and some instances of glassiness. A change in the short-time dynamics of the system allows it to enter a metastable state, which would have been otherwise avoided and whose lifetime can easily become exceptionally long at (experimentally relevant) low temperatures.
\begin{figure}[t]
  \centering
  \includegraphics[width=\linewidth]{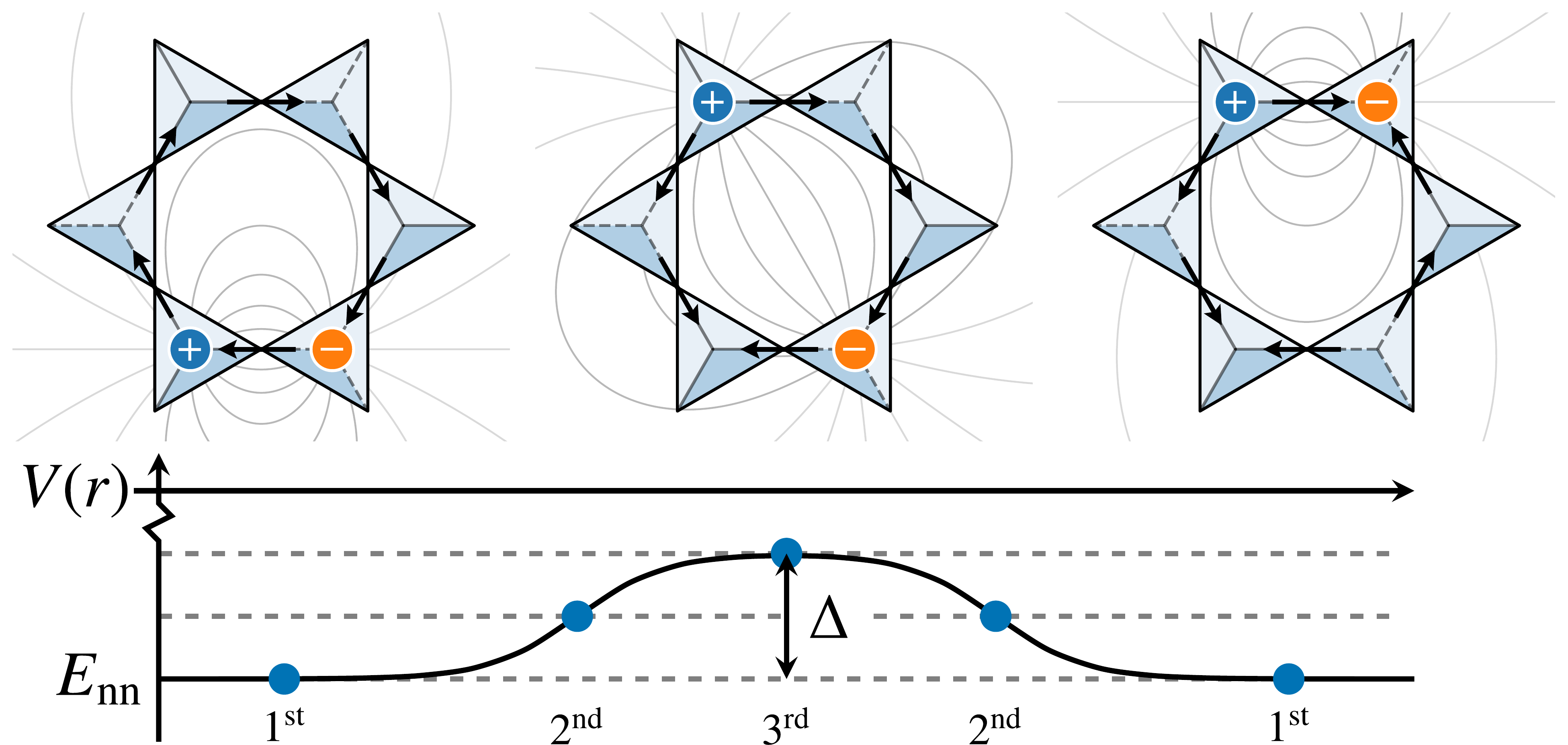}
  \caption{Schematic depiction of a noncontractible monopole--antimonopole pair, responsible for the metastable plateau in monopole density observed following a thermal quench to low temperatures in classical spin ice. The activated decay of the pair requires separating its members up to third-neighbour distance, as shown in the central figure, costing an energy $\Delta$ (in isolation) due to their mutual Coulombic attraction. The pair is then able to annihilate elsewhere on the lattice, as shown for example in the rightmost figure.}
  \label{fig:nc-pair}
\end{figure}

Our results further demonstrate that the plateau reduces to a finite size effect when considering a model with solely finite-range interactions between monopoles.
Hence, the experimental observation of a metastable plateau corresponding to a finite density of noncontractible pairs in spin ice is direct evidence of the long-range nature of the interactions between the monopoles.
This adds one important experimental avenue to study these interactions, whose range has thus far been probed only via the field-dependence of unbinding of monopole pairs~\cite{Paulsen2016}, and indirectly via the appearance of a liquid-gas phase diagram~\cite{Castelnovo2008}.

Our findings are particularly timely thanks to the recent experimental claim that a state rich in noncontractible pairs can be generated in classical spin ice materials {\DTO} and {\HTO}~\cite{Paulsen2016} using a so-called avalanche quench protocol~\cite{Paulsen2014}.

The paper is structured as follows. We start by reviewing the background on thermal quenches in classical spin ice and by summarising the main results obtained in this work in Sec.~\ref{sec: background and summary}. We then provide an overview of the models we consider in Sec.~\ref{sec:models}, and we present our Monte Carlo results in Sec.~\ref{sec:numerics}, including a finite size scaling analysis of the density of noncontractible pairs in the metastable plateau.
Section~\ref{sec:meanfield} is devoted to the use of mean field population dynamics to understand the differences in behaviour between the various models and types of interaction. We draw our conclusions and highlight the relevance of our results to experiments in Sec.~\ref{sec:conclusions}.


\section{\label{sec: background and summary}
Background and summary of results
        }

Dipolar spin ice systems have been predicted to exhibit dynamically-arrested, monopole-rich, metastable states following appropriate thermal and field quenches~\cite{Castelnovo2010,Mostame2014}.
Reference~\onlinecite{Castelnovo2010} recognised that at the heart of the dynamical arrest lies the formation of so-called noncontractible pairs: a monopole and an antimonopole sitting next to one another, separated by a spin whose reversal does not lead to their annihilation. As a result, the two defects become bound to one another and are unable to move throughout the lattice without separating---a process that costs Coulomb energy due to the mutual attraction between the two opposite charges~\footnote{Noncontractible pairs cannot move from the site upon which they form without separating to third neighbour distance. Otherwise, movement of the pair would require motion of a monopole along a blocked direction. Indeed, the caterpillar-like motion of separating and rejoining, trailing one another, is prevented by the intervening spin being a blocked direction for the trailing monopole.}. This activation energy barrier explains why a noncontractible pair per se is metastable.

\begin{figure}[t]
  \centering
  \subfloat{\includegraphics[width=0.333\linewidth,valign=c]{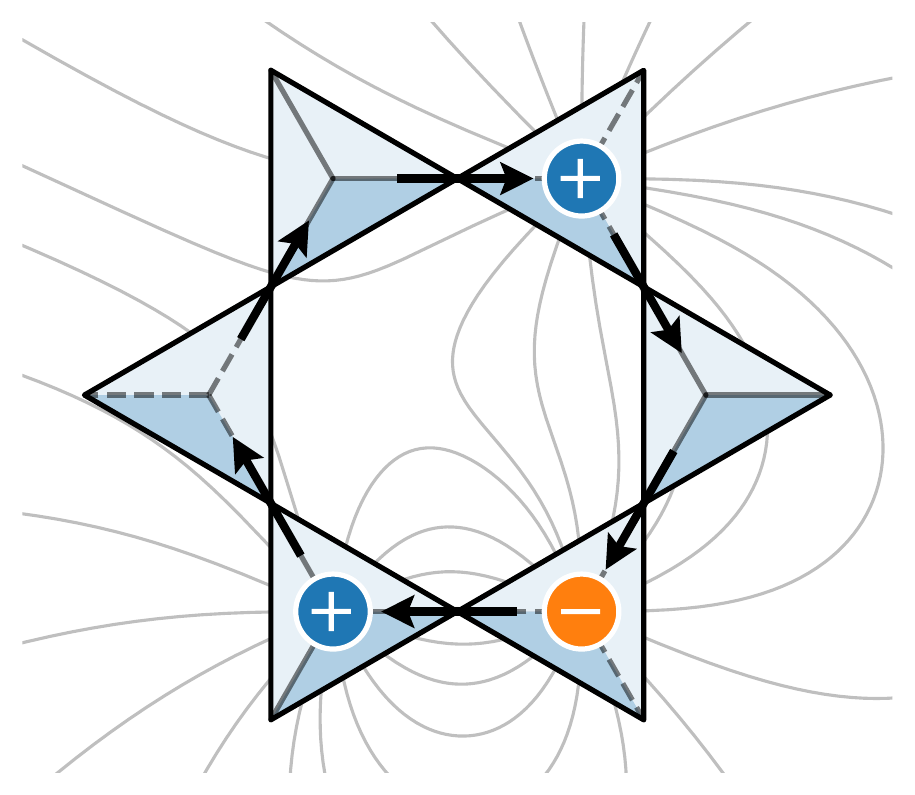}}
  \subfloat{\includegraphics[width=0.333\linewidth,valign=c]{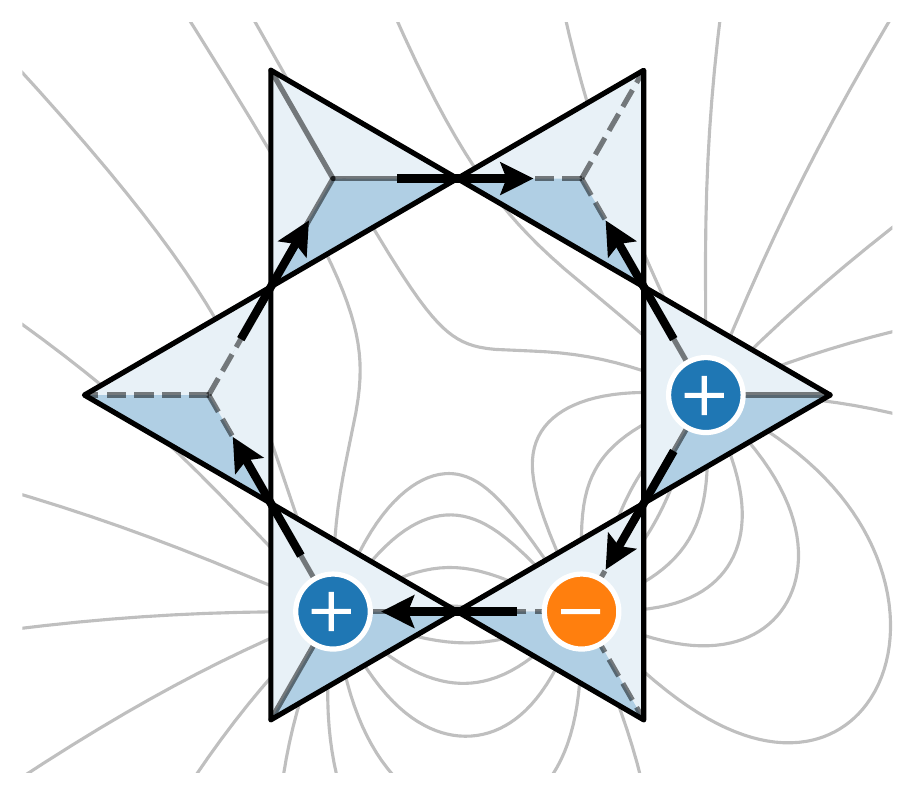}}
  \subfloat{\includegraphics[width=0.333\linewidth,valign=c]{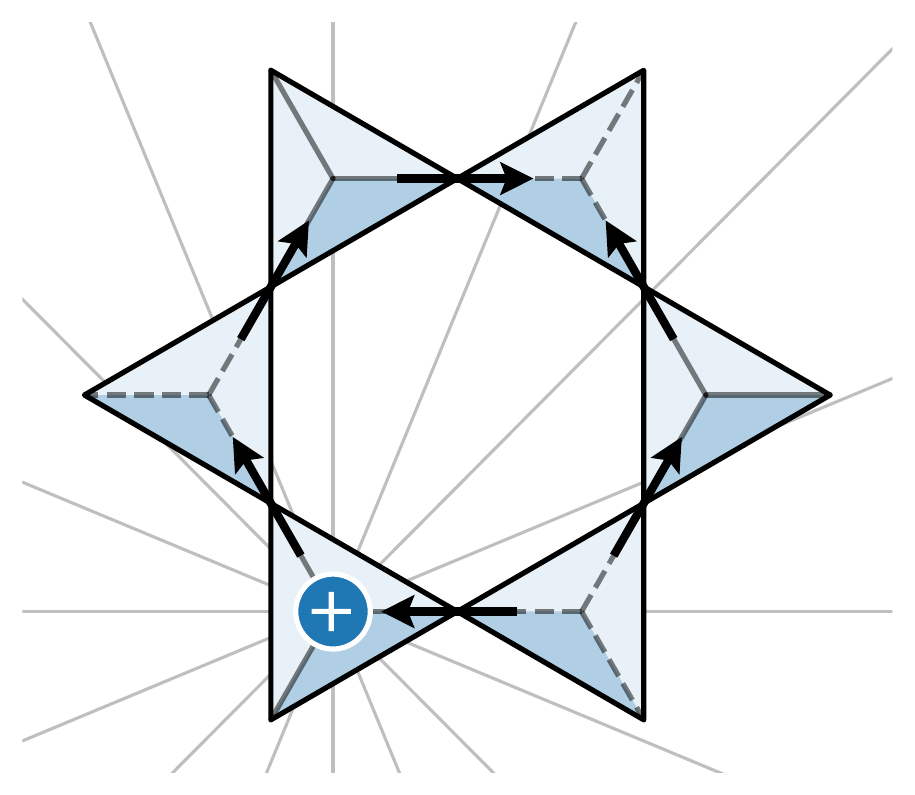}}
  \caption{Schematic depiction of monopole-assisted decay of a noncontractible pair. A free monopole annihilates with the oppositely charged member of the stationary noncontractible pair, thereby freeing up its partner. All moves shown lower the energy of the system and hence monopole-assisted decay is the dominant decay avenue for noncontractible pairs with respect to thermally activated decay when free monopoles are abundant in the system.}
  \label{fig:monopole-assisted}
\end{figure}

In general, two decay channels are available to noncontractible pairs. Firstly, they can separate and annihilate somewhere else on the lattice at the cost of paying an activation energy barrier; the smallest barrier associated with such activated decay processes requires separating the pair up to third-neighbour distance, as shown in Fig.~\ref{fig:nc-pair}. Alternatively, pairs can undergo monopole-assisted decay: When the pair is hit by a stray (free) monopole, this causes the annihilation of the oppositely charged member of the pair, thus freeing up its partner~\cite{Castelnovo2010}, as in Fig.~\ref{fig:monopole-assisted}.
This second process does not incur an energy barrier and does not change the density of free monopoles.

In equilibrium, a useful quasiparticle description for spin ice is in terms of deconfined magnetic charges~\cite{Castelnovo2008}. Conversely, the long (intrinsic) lifetime of noncontractible pairs justifies their introduction as an effectively distinct ``species'' of quasiparticle when studying classical spin ice in the strongly nonequilibrium setting of thermal and field quenches, as demonstrated already in Refs.~\onlinecite{Castelnovo2010,Mostame2014}.

The mere existence of noncontractible pairs in the system however does not warrant the appearance of a macroscopic metastable state. Indeed, when free monopoles are abundant, non-activated (fast) monopole-assisted decay is the leading relaxation channel with respect to thermally-activated (slow) decay of noncontractible pairs, and one does not expect any metastable plateau to appear. It is only when the system undergoes a ``population inversion'' (in contrast to thermodynamic equilibrium), where noncontractible pairs become the dominant species with respect to free monopoles, that the activation energy barrier to decay can induce a long-lived metastable plateau at low temperatures.
This is indeed what one observes in numerical simulations of dipolar spin ice, following appropriate thermal quenches~\cite{Castelnovo2010}.

The aforementioned population inversion is key to the metastable plateau. Its origin however
was not investigated in Ref.~\onlinecite{Castelnovo2010} and is the subject of the present work.
We find that it ultimately rests on the long-range tail of the Coulomb interaction between monopoles.
This can be qualitatively understood as being due to the energetic bias in the motion of monopoles in the far field. Monopole--antimonopole collision events are subject to a Coulombic charge--charge attraction ($\propto \! r^{-2}$), whereas collisions between a free monopole and a noncontractible pair are subject to weaker charge--dipole interactions ($\propto \! r^{-3}$).
This leads to a bias that increases the likelihood of free monopoles annihilating (or forming new noncontractible pairs) over their chance of annihilating existing noncontractible pairs via monopole-assisted decay.
Further, since the final temperature in the thermal quenches is much less than all other energy scales in the problem, the system enters a non-hydrodynamic regime where the monopoles move at terminal velocity in the direction of the local force acting on them.
This allows the system to violate the law of formal kinetics~\cite{Ovchinnikov2000} and to exhibit a decay of the free monopole density faster than inverse time.
The combination of the long-range bias and `terminal velocity' motion of free charges leads to a rapid decay of the free monopole density in the system, leaving behind an excess of noncontractible pairs.
This is ultimately the linchpin of the finite-density metastable plateau observed in numerical simulations.

This behaviour is most remarkable. By altering the dynamics of what is ultimately a transient regime, spin ice is able to enter a metastable state whose lifetime for experimentally relevant temperatures and system sizes may well exceed any realistically accessible timescales (of order one year in Fig.~\ref{fig: Coulomb vs dipolar} when expressed in physical units).

We verify this scenario through extensive numerical Monte Carlo simulations of thermal quenches in spin ice systems with nearest-neighbour spin--spin interactions and long-range Ewald-summed magnetic Coulomb interactions between defective tetrahedra~\cite{Jaubert2009}. Upon truncating the Coulomb interactions to finite range, the long-range bias is removed. We find that the finite-density metastable plateau correspondingly disappears in the thermodynamic limit.
These findings are corroborated (in Sec.~\ref{sec:MC:charges}) by directly simulating mutually interacting magnetic charges hopping on a diamond lattice (with no Dirac strings), for which we observe qualitatively similar behaviour.

To supplement the numerics, we provide an analytical understanding of both the value of the plateau in the thermodynamic limit and its finite size scaling using mean field population dynamics, treating the system as a Coulomb liquid of magnetic charges. We show how the ratio of the rate of monopole-assisted decay to the rate of charge--charge annihilation underpins both the finite size scaling exponent in the case of truncated interactions, and the density at which the plateau occurs in the long-range case.


\section{\label{sec:models}
Models}

In this work, we contrast the effect of truncating the Coulomb interactions between monopoles in spin ice with the same truncation in a system of magnetic charges hopping on a diamond lattice.
The latter model is defined without reference to any underlying spin configuration, i.e., without Dirac strings connecting opposite charges, which allows us further clarity in ascertaining their role
in thermal quenches and the formation of the metastable plateau.


\subsection{\label{sec:models:CSI}
Classical spin ice}

The canonical model of classical spin ice (CSI) consists of exchange ($J$) and dipolar ($D$) interactions between classical Ising spins $\v{S}_i$, which live on the sites of a pyrochlore lattice~\cite{Siddharthan1999, Hertog2000}.
The crystal field anisotropy in spin ice materials (e.g., {\DTO} and {\HTO}) constrains the spins to point along the local $[111]$ directions, $\v{e}_i$. Absorbing the magnitude of the spins into the definition of the coupling constants, we can therefore represent them as $\v{S}_i = S_i \v{e}_i$, with $S_i \in \{-1, +1\}$, and write the dipolar spin ice Hamiltonian as
\begin{eqnarray}
  H_d(\{S_i\}) &=&
    \frac{J}{3} \sum_{\langle ij \rangle} S_i S_j
	\\
	&+& D \sum_{(ij)} \left[ \frac{\v{e}_i\!\cdot\!\v{e}_j}{|\v{r}_{ij}|^3} - \frac{3(\v{e}_i\!\cdot\!\v{r}_{ij})(\v{e}_j\!\cdot\!\v{r}_{ij})}{|\v{r}_{ij}|^5} \right] S_i S_j
  \, ,
\nonumber
\end{eqnarray}
where, in the first line, we used the fact that $\v{e}_i \cdot \v{e}_j = -1/3$ for any nearest neighbour pair of sites, $\langle ij \rangle$.

For the majority of this work, we use an effective Hamiltonian in which the exchange and dipolar interactions between the spins are retained only at nearest-neighbour level, and farther range couplings are accounted for effectively by a pairwise interaction $V(\{Q_a\})$ between tetrahedral charges $Q_a$,
\begin{equation}
  H_c(\{ S_i \}) = J_\text{eff}\sum_{\langle i j \rangle} S_i S_j + E_\text{nn} \sum_{a < b} \frac{Q_a Q_b}{r_{ab}}
  \, ,
  \label{eq: long range case}
\end{equation}
where $i, j$ index the sites of the pyrochlore lattice, $a, b$ index the tetrahedra and $r_{ab} = |\v{r}_a - \v{r}_b|/r_\text{nn}$ is the distance between the centres of tetrahedra $a$ and $b$ in units of the distance between neighbouring tetrahedra.
The charge on tetrahedron $a$ is $Q_a = \pm \sum_{i \in a} S_i/2$, where the sign depends on the sublattice that $a$ belongs to. The charges $Q_a$ therefore assume the values $Q_a \in \{ 0, \pm 1, \pm 2 \}$, where $Q_a = \pm 1$ are dubbed monopoles and $Q_a = \pm 2$ double monopoles.
We use the convention that a positive charge corresponds to a majority of spins pointing out of a given tetrahedron.
Two equally charged monopoles on neighbouring sites have a Coulomb energy $E_\text{nn}$ (in an infinite system).
Throughout the manuscript we use an effective exchange coupling $J_\text{eff}=\SI{1.463}{\kelvin}$
\footnote{This value of the effective exchange coupling was obtained using the chemical potential $\mu = -\SI{8.92}{\kelvin}$ in Ref.~\cite{Jaubert2011}. In particular, we use $\mu = -4J_\text{eff} - E_\text{nn}$ to define $J_\text{eff}$, i.e., (minus) the energy required to create a pair of oppositely charged monopoles and separate them to infinity.
The dynamics of the system is however not particularly sensitive to the precise value of $J_\text{eff}$, as long as the ground state remains unchanged.}
and nearest-neighbour Coulomb energy $E_\text{nn} = \sqrt{128/27}D = \SI{3.06}{\kelvin}$, appropriate for the classical spin ice compound \DTO.
Such an effective description~\eqref{eq: long range case} is quantitatively accurate, up to quadrupolar corrections, by virtue of projective equivalence~\cite{Isakov2005} (and this is indeed the case also in thermal quenches, as illustrated in Fig.~\ref{fig:nc-pair-density}).
With these parameters, the macroscopically degenerate ground state manifold corresponds to the charge vacuum, $Q_a=0$, $\forall a$, i.e., a $2\,$in-$2\,$out configuration of spins on each tetrahedron.

We note that the nearest-neighbour exchange interaction between spins can be viewed as a chemical potential of size $2J_\text{eff}$ for the monopoles (namely, the charges $Q_a = \pm 1$):
\begin{equation}
  J_\text{eff}\sum_{\langle i j \rangle} S_i S_j = 2 J_\text{eff} \sum_a Q_a^2 - N_s J_\text{eff}
	\, .
\end{equation}
This interpretation however no longer holds straightforwardly in the presence of double monopoles.

To test the role of the long-range tail of the Coulomb interaction in the appearance of the population inversion, we also consider a similar model where the interactions $V(\{ Q_a \})$ between monopoles are truncated at nearest-neighbour distance:
\begin{equation}
  H_t(\{ S_i \}) = J_\text{eff} \sum_{\langle i j \rangle} S_i S_j + \Delta \sum_{\langle a b \rangle } Q_a Q_b
  \, .
  \label{eqn:CSI-truncated-Hamiltonian}
\end{equation}
This model will be referred to as classical spin ice with truncated
interactions.
Such a nearest-neighbour interaction between monopoles allows for the formation of noncontractible pairs without inducing any long-range energetic bias in the motion of the monopoles.

Separating an isolated pair of nearest-neighbour monopoles with charge $Q=\pm 1$ in this model costs an energy $\Delta$.
To preserve the behaviour of the system (primarily its ground state), the truncation of the interactions must be done with care.
We choose the value of $\Delta$ such that the energy barrier to separating a noncontractible pair around a hexagonal plaquette (as depicted in Fig.~\ref{fig:nc-pair}) is equal in the cases of truncated~\eqref{eqn:CSI-truncated-Hamiltonian} and long-ranged~\eqref{eq: long range case} interactions~\footnote{This energy barrier is equal to the Coulomb energy required to separate the pair to third neighbour distance, $r_\text{3n}$, i.e., $\Delta = E_\text{nn}(1 - r_\text{nn}/r_\text{3n})$.}:
\begin{equation}
  \Delta = E_\text{nn} \left( 1 - \sqrt{\frac{3}{11}} \right) \simeq \SI{1.46}{\kelvin}
  \, .
\end{equation}
Such a choice preserves the charge vacuum ground state, and since the energy barrier for the activated decay of noncontractible pairs is equal for both types of interaction, the demise of a possible metastable plateau will occur at similar times in the two cases.

The difference between the single spin flip dynamics of the two Hamiltonians, $H_c$ and $H_t$, therefore rests solely in the long-range energetic bias in the motion of monopoles across the system.
In a finite system containing $L^3$ cubic unit cells, the total number of spins is $N_s=16L^3$, and the number of tetrahedra is $N_t = 8L^3$.
In our simulations, we use periodic boundary conditions and we deal with long-ranged interactions (dipolar as well as Coulomb) using the method of Ewald summation~\cite{deLeeuw1980,Frenkel2001}.

We note that there also exists a long-range Coulomb interaction between monopoles of entropic origin~\cite{Henley2010}.
As we are unable to alter the range of the entropic interactions, we introduce in the following section a family of charge models that live on the diamond lattice
in which the charges are not born out of underlying spin configurations.
This will allow us to observe that the role of entropic interactions in thermal quenches is in fact negligible and hence they will not be discussed further in our work. This is shown most directly by the good quantitative agreement between the classical spin ice and charge model simulations, and the mean field analytics, for truncated interactions.
%
%

\subsection{\label{sec:models:charges}
Charges on diamond lattice}

\begin{figure}[t]
\includegraphics[width=\columnwidth]{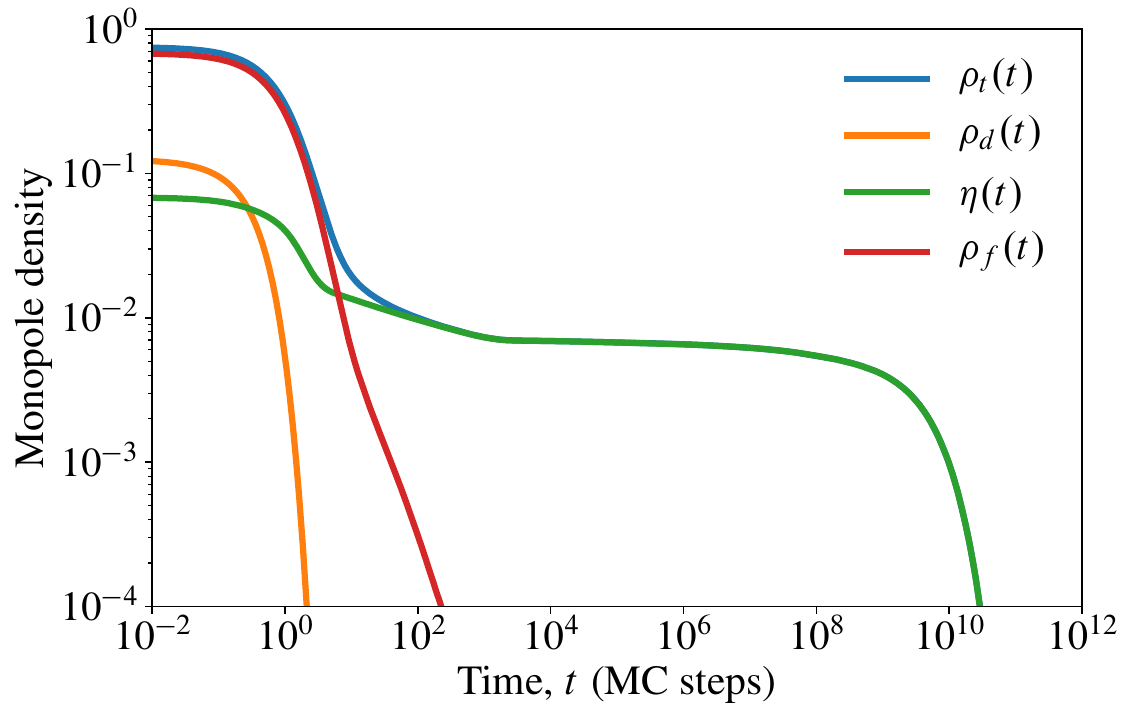}
\caption{\label{fig: Coulomb vs dipolar}
Monte Carlo simulations of a thermal quench in spin ice subject to Ewald-summed Coulomb interactions between monopoles [Hamiltonian~\eqref{eq: long range case}, system size $L=22$, i.e., $\num{170368}$ spins] from infinite temperature down to $T = \SI{0.06}{\kelvin}$. The curves show the evolution of the averaged total density of monopoles per tetrahedron $\rho_t$ (blue), the free monopole density $\rho_f$ (red), the density of monopoles forming noncontractible pairs $\eta$ (green) and the double charge density $\rho_d$ (orange). Time is expressed in units of Monte Carlo steps per site, and the densities are averaged over $\num{4096}$ histories.}
\end{figure}
\begin{figure}[t]
\includegraphics[width=\columnwidth]{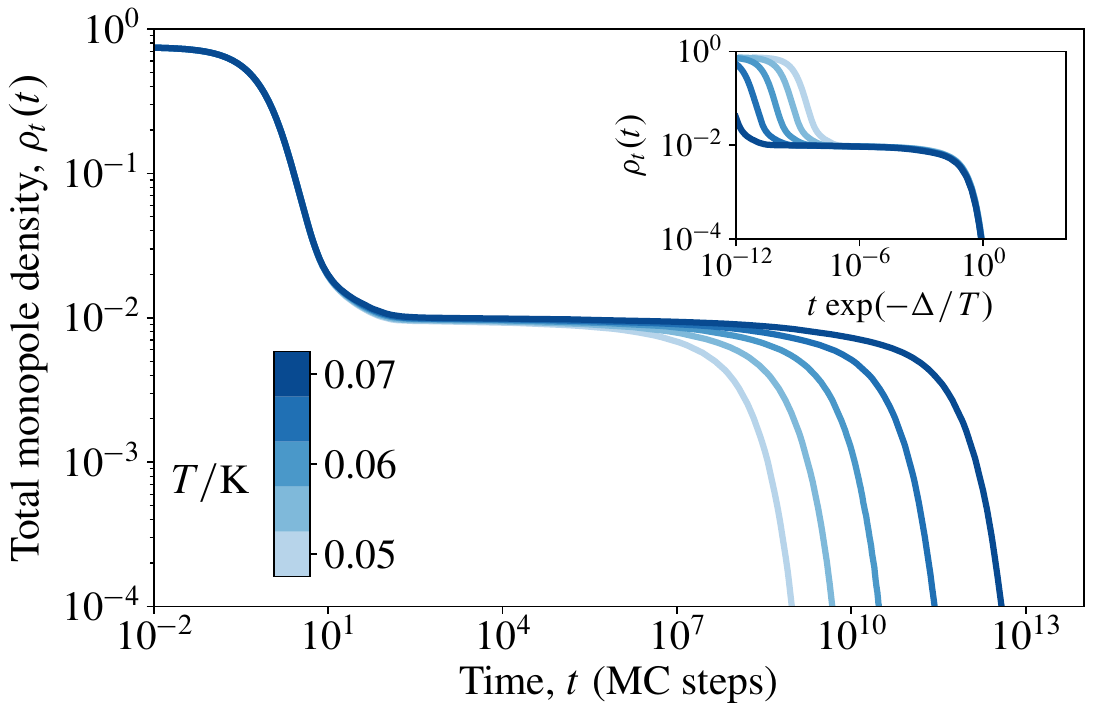}
\caption{\label{fig: LongScaled}
Monte Carlo simulations of the total density of monopoles $\rho_t$ in spin ice in the case of long-range Coulomb interactions between monopoles, Eq.~\eqref{eq: long range case}, after a thermal quench from infinite temperature down to final temperatures $T = 0.05$--$\SI{0.07}{\kelvin}$ (in equidistant steps) for a system of size $L=8$, i.e., $\num{8192}$ spins. The densities are averaged over $\num{4096}$ histories.
Inset: the same curves plotted after rescaling the time axis by a factor $\exp(\Delta/T)$, where $\Delta \simeq \SI{1.46}{\kelvin}$ is the Coulomb energy barrier incurred by separating two monopoles around a hexagonal plaquette, showing an excellent collapse of the long-time decay of the monopole density.
}
\end{figure}

To identify the role of the spin configuration underlying each monopole configuration, we also consider two further effective models of charges $Q_a$ hopping on a diamond lattice, thereby removing any entropic effects and blocked directions associated with the spins (in particular, by removing the underlying spin network, there are no Dirac strings associated with the magnetic charges in the following models).
We restrict our simulations to the relevant charge values $Q_a \in \{ 0,\pm 1,\pm 2\}$ only.
These charge models (CM) also allow for a more direct comparison with our analytical mean field modelling (see Sec.~\ref{sec:meanfield}), which largely neglects the aforementioned complications associated with the spinful description of the system's dynamics.

In the case of long-range interactions between the charges, we use the Hamiltonian
\begin{equation}
  H^{\rm CM}_c(\{Q_a\}) = 2J_\text{eff} \sum_a Q_a^2 + E_\text{nn} \sum_{a < b} \frac{Q_a Q_b}{r_{ab}}
  \, ,
  \label{eqn:charges-hamiltonian-long-range}
\end{equation}
subject to the hard constraint that each site may not be occupied by more than two charges. This model will be referred to as the long-range interacting
charge model.

The Hamiltonian~\eqref{eqn:charges-hamiltonian-long-range} must be further supplemented by rules which govern the dynamics of the charges.
Namely, in order to take into account the effect of noncontractible pairs, when two opposite (single) charges come into nearest-neighbour contact, there exists some finite probability, $p_\text{nc}$, of forming a noncontractible pair. If a noncontractible pair is formed, it is then not possible for the charges to annihilate along their common bond. At finite temperature, their activated decay can be accounted for by associating an energy barrier $\Delta$ with this process.

The probability $p_\text{nc}$ can be estimated by counting the number of spin configurations compatible with two oppositely charged monopoles on adjacent tetrahedra, and taking the fraction thereof that correspond to a noncontractible pair. Considering the minimal cluster of two tetrahedra only ($7$ spins in total), one finds that the relevant fraction is $p_\text{nc} = 1/10$~\cite{Castelnovo2010}. Extending the calculation to larger clusters does not lead to significant variation in this value; for example, considering a full hexagon of tetrahedra involving the two monopoles gives $p_\text{nc} = 41/406$.
Further, small perturbations in $p_\text{nc}$ do not appreciably modify the dynamics of the system.

For the case of truncated interactions between the tetrahedral charges, the Hamiltonian becomes
\begin{equation}
  H^{\rm CM}_t(\{Q_a\}) = 2J_\text{eff} \sum_a Q_a^2 + \Delta \sum_{\langle a b \rangle} Q_a Q_b
  \, ,
  \label{eqn:charges-hamiltonian-truncated}
\end{equation}
referred to as the charge model with truncated interactions.
The model is again subjected to the same constraints on charge values and dynamics.
The difference between the two charge models, $H^{\rm CM}_c$ and $H^{\rm CM}_t$, lies only in the long-range energetic bias associated with the Coulomb interaction.


\section{\label{sec:numerics}
Monte Carlo Simulations}


\subsection{\label{sec:numerics:CSI}Classical spin ice}
\subsubsection{Long-range Coulomb interactions}

%
%
\begin{figure}[t]
\includegraphics[width=\columnwidth]{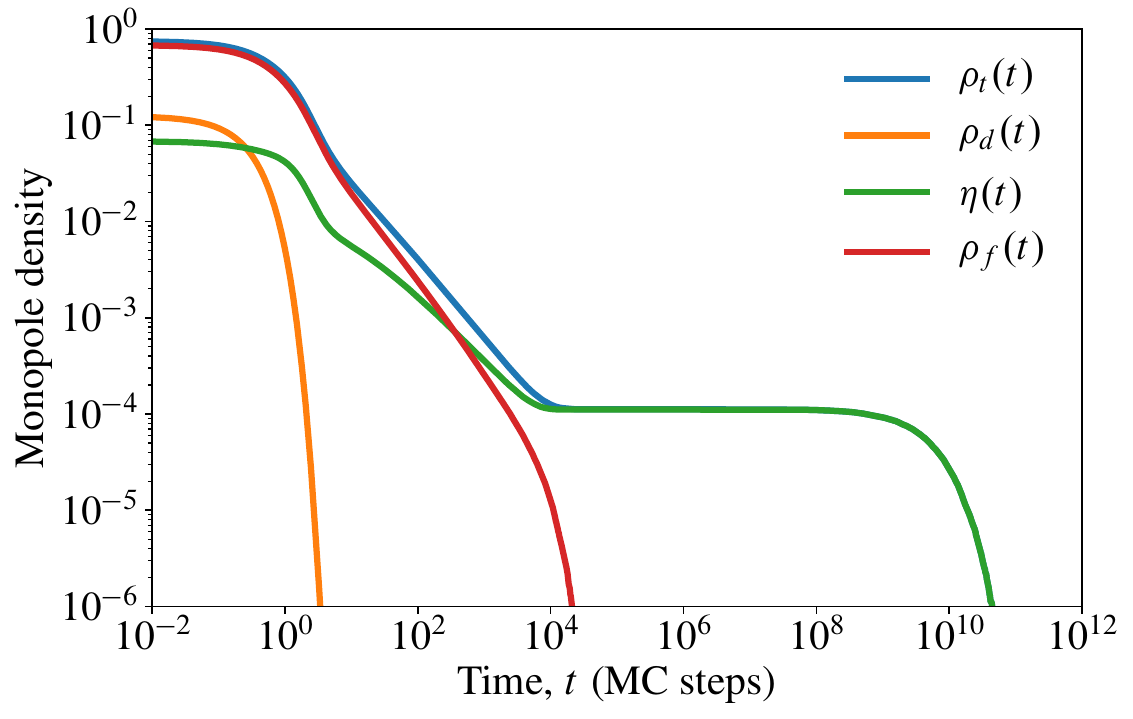}
\caption{\label{fig: truncated Coulomb}
Monte Carlo simulations of a thermal quench in spin ice where the interactions between monopoles are truncated to nearest-neighbour distance [Hamiltonian~\eqref{eqn:CSI-truncated-Hamiltonian}, system size $L=16$, i.e., $\num{65536}$ spins] from infinite temperature down to $T = \SI{0.06}{\kelvin}$. Time is expressed in units of Monte Carlo steps per site, and the densities are averaged over $\num{4096}$ histories. The metastable plateau due to noncontractible pairs of monopoles remains present, but occurs at lower densities and at later times than in the case of long-range interactions (cf.~Fig.~\ref{fig: Coulomb vs dipolar}).
}
\end{figure}
%
%
%
\begin{figure}[t]
\includegraphics[width=\columnwidth]{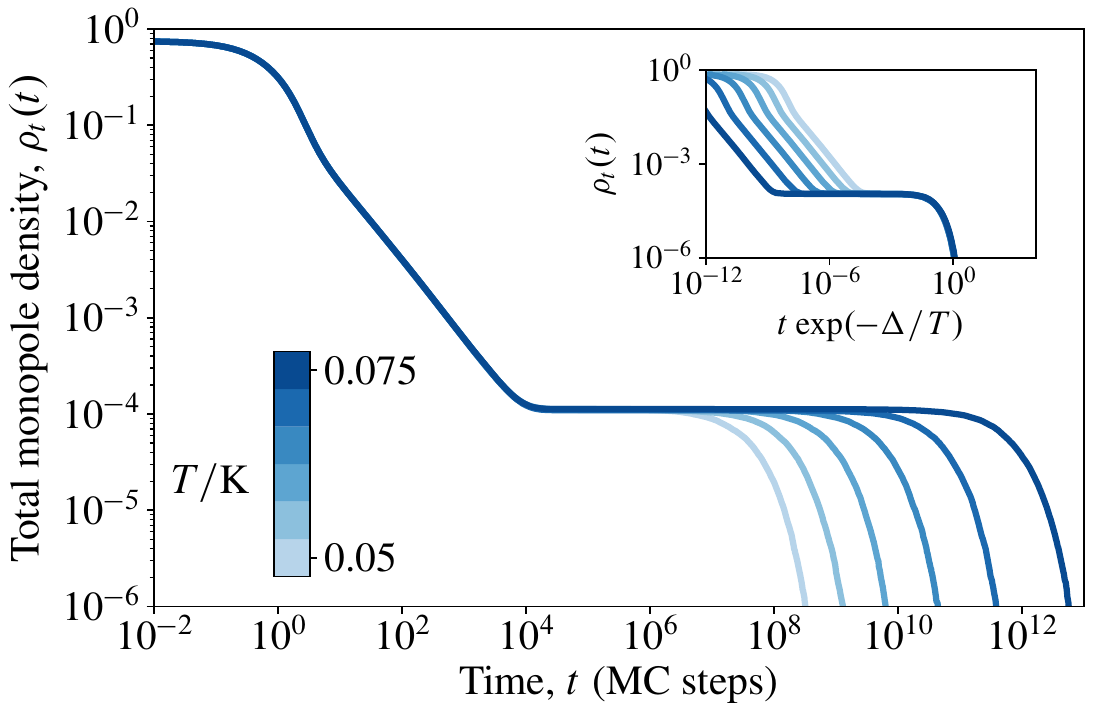}
\caption{\label{fig: TruncScale}
Monte Carlo simulations of the total density of monopoles $\rho_t$ in spin ice in the case of truncated interactions between monopoles, Eq.~\eqref{eqn:CSI-truncated-Hamiltonian}, after a thermal quench from infinite temperature down to various temperatures $T = 0.05$--$\SI{0.075}{\kelvin}$ (in equidistant steps) for a system of size $L=16$, i.e., $\num{65536}$ spins. The densities are averaged over $4096$ histories. Inset: the same curves plotted after rescaling the time axis by a factor $\exp(\Delta/T)$, showing an excellent collapse of the long-time decay.
}
\end{figure}

In Fig.~\ref{fig: Coulomb vs dipolar} we show the monopole density evolution following a thermal quench, as in Ref.~\onlinecite{Castelnovo2010}, simulated using the modified Monte Carlo code, corresponding to~\eqref{eq: long range case}, instead of the conventional dipolar Monte Carlo (for a direct comparison, see Fig.~\ref{fig:nc-pair-density}).
We use single spin flip dynamics and the Waiting Time Method~\cite{Dall2001, Dall2003} to access long simulation times at low temperatures (see Appendix~\ref{app:simulation-details} for some details specific to our simulations). The system is initially prepared in the paramagnetic phase at infinite temperature, then at  $t=0$ the temperature is set to its target value, $T \ll J_\text{eff}$, and we start measuring various monopole densities as a function of time~\footnote{In {\DTO} these initial conditions are experimentally relevant to initial temperatures $T \gg \SI{1}{\kelvin}$, and the quench in temperature should occur over timescales less than $\sim \SI{1}{\milli\second}$, the characteristic single spin flip timescale~\cite{Snyder2004}.}. These densities are then averaged over many histories with different random initial conditions sampled from the infinite temperature ensemble. We find good agreement with the dynamical arrest observed in Ref.~\onlinecite{Castelnovo2010}:
Rather than rapidly equilibrating to a monopole-sparse state, we observe instead the emergence of a metastable plateau in the monopole density due to noncontractible monopole--antimonopole pairs.

Specifically, we measure the total monopole density (monopoles per tetrahedron) in the system, $\rho_t$, counting all-in and all-out tetrahedra as doubly occupied sites; the fraction of such doubly occupied sites, $\rho_d$; the density of monopoles forming noncontractible pairs, $\eta$; and the `free' monopole density~\footnote{Note that there are many possible definitions of the `free' monopole density due to ambiguities that arise in defining pairs of monopoles in the monopole-dense (short-time) limit. However, all definitions agree once the typical separation of monopoles is greater than $r_{\text{nn}}$.} $\rho_f \equiv \rho_t - \eta$, i.e., the density of monopoles that do not form noncontractible pairs. A noncontractible pair is \emph{defined} as a pair of adjacent, oppositely-charged monopoles for which the reversal of the intervening spin shared by the two tetrahedra does not lead to annihilation of the pair.

In isolation, the barrier to activated decay of a noncontractible pair is $\Delta \simeq \SI{1.46}{\kelvin}$.
In the presence of a finite density $\eta$ of other noncontractible pairs, the distribution of energy barriers is broadened around a mean value of $\Delta$ due to dipole--dipole interactions between the pairs.
Given that the Coulombic approximation to the monopole--monopole interaction neglects quadrupolar corrections, we expect the distribution of such energy barriers to be more sharply peaked than in the dipolar case. This is indeed confirmed by the excellent collapse of the long-time decay of the total monopole density for various temperatures upon rescaling the time axis by a factor $\exp(\Delta/T)$, as illustrated in Fig.~\ref{fig: LongScaled} (see also Fig.~\ref{fig:nc-pair-density}, where the dipolar case shows a correspondingly broader decay of the metastable plateau).


\subsubsection{Truncated interactions}

In Fig.~\ref{fig: truncated Coulomb} we plot the various monopole densities for an identical thermal quench for the case of truncated interactions between monopoles in classical spin ice [i.e., Eq.~\eqref{eqn:CSI-truncated-Hamiltonian}]. A metastable plateau remains present in the dynamics of the system, and once again the behaviour of the monopole densities tells us that it is clearly due to noncontractible pairs. The plateau however occurs at substantially lower densities and the onset occurs at later times when compared with the corresponding long-range interacting system, Eq.~\eqref{eq: long range case}, in Fig.~\ref{fig: Coulomb vs dipolar}.

The decay of the monopole density at long times collapses for a range of temperatures upon rescaling the time axis by a Boltzmann factor $\exp(\Delta/T)$, as illustrated in the inset of Fig.~\ref{fig: TruncScale}, confirming that the thermally activated decay of noncontractible pairs is again responsible for the eventual demise of the plateau at a time $\tau_\text{nc} \sim \exp(\Delta/T)$. Once a given pair has separated, the two constituent monopoles may find each other and annihilate by performing a random walk, the shortest of which is around a single hexagonal plaquette.
Since the noncontractible pairs do not interact beyond a fixed, finite separation, the energy barriers are $\delta$-distributed about $\Delta$.

%

\subsubsection{\label{sec:CSI:finite-size-scaling}
Comparison and finite size scaling
              }

\begin{figure}[t]
\includegraphics[width=\columnwidth]{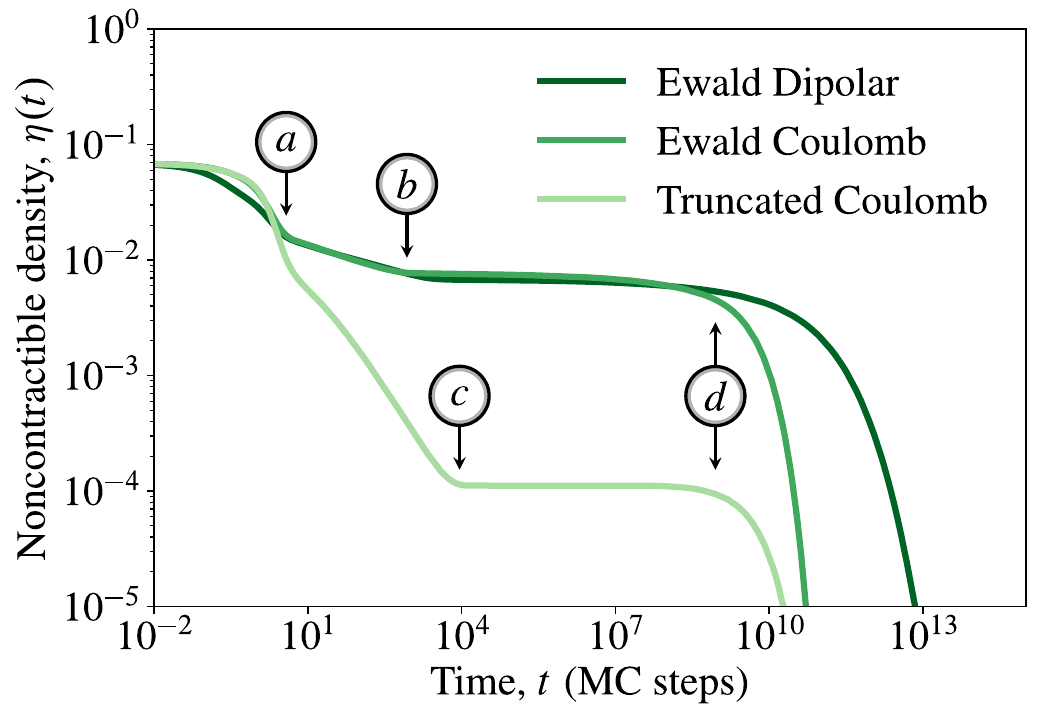}
\caption{Comparison of noncontractible pair densities $\eta(t)$ for the three types of interaction for a thermal quench from infinite temperature down to $T = \SI{0.06}{\kelvin}$ (system size $L=16$, i.e., $\num{65536}$ spins) in classical spin ice. Time is expressed in units of Monte Carlo steps per site, and the densities are averaged over $\num{4096}$ histories. The markers labelled $a$, $b$, $c$, and $d$ identify the boundaries between the four dynamical regimes discussed in the main text. At ($a$), nearly all doubly occupied sites have been removed from the system. Points ($b$) and ($c$) mark the onset of the metastable plateau for the cases of long-range and truncated interactions, respectively.
At ($d$), the noncontractible pairs decay via thermal activation.
}
\label{fig:nc-pair-density}
\end{figure}

\begin{figure}[t]
\includegraphics[width=\columnwidth]{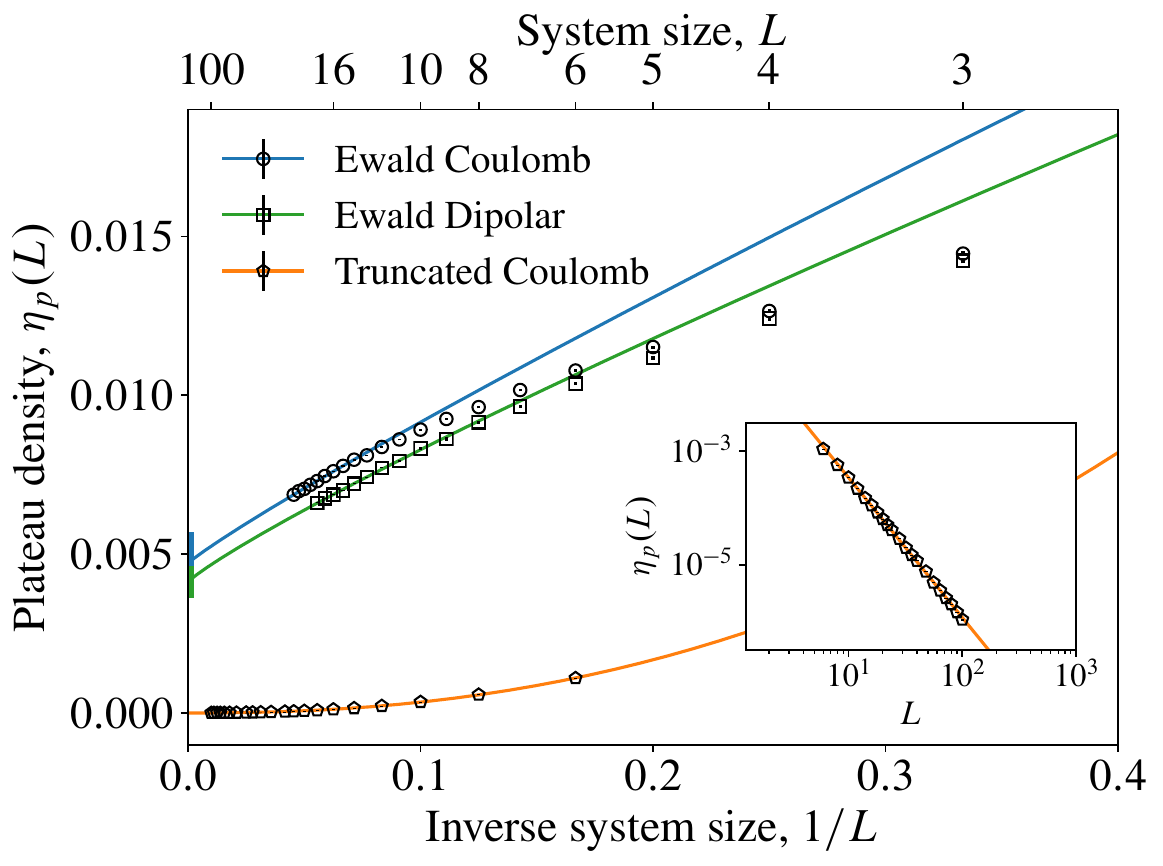}
\caption{Finite size scaling of the plateau in noncontractible monopole density $\eta_p(L)$ for long-range Coulomb and truncated (nearest-neighbour) interactions between monopoles, and long-range dipolar interactions between spins in classical spin ice. The data are averaged over at least $\num{4096}$ histories. The lines are fits to the scaling ansatz $\eta_p(L) - \eta_p(\infty) \sim L^{-\nu}$, while the symbols represent the Monte Carlo data. The corresponding error bars are smaller than the width of the fit lines. In the truncated case (system sizes $L=6$--$100$ inclusive), the data are consistent with a plateau that vanishes in the thermodynamic limit. This is verified using a log--log plot of the plateau density against system size $L$ in the inset. Conversely, the long-range Coulomb ($L=3$--$22$ inclusive) and dipolar ($L=3$--$18$ inclusive) cases appear to exhibit a nonvanishing noncontractible pair density in the metastable plateau in the thermodynamic limit: $\eta_p(\infty) = 4.7(9) \times 10^{-3}$ and $\eta_p(\infty) = 4.1(5) \times 10^{-3}$, respectively.
}
\label{fig: plateau scaling}
\end{figure}

In Fig.~\ref{fig:nc-pair-density} we plot the noncontractible pair density as a function of time, $\eta(t)$, for all three types of interaction introduced in Sec.~\ref{sec:models:CSI} for classical spin ice: Ewald-summed dipolar interactions between spins, Ewald-summed Coulomb interactions between monopoles, and truncated (nearest-neighbour) interactions between monopoles.

In each of the three cases, the time evolution of $\eta(t)$ can be decomposed into four dynamical regimes.
The crossovers between these regimes are identified by the markers (\textit{a})--(\textit{d}) in Fig.~\ref{fig:nc-pair-density}.
\begin{enumerate}
\item At times before (\textit{a}), the large exchange energy cost associated with double occupancy of a tetrahedron ($8J_\text{eff}$) and the ability of such configurations to decay spontaneously ensures that doubly occupied sites are removed rapidly during this regime (exponentially fast in time, see Appendix~\ref{app: double charges} for details).

\item Between (\textit{a})--(\textit{b}) and (\textit{a})--(\textit{c}) the differences between the three interaction types become manifest.
In all cases we observe a much slower decay of the noncontractible pair density once the double monopoles have been removed from the system.
However, the rate of decay and the timescales over which this decay occurs are vastly different for the truncated [(\textit{a})--(\textit{c})] versus long-range interacting [(\textit{a})--(\textit{b})] models.
In the Coulomb and dipolar cases, the long-range nature of the interactions leads to an energetic bias which favours monopole--antimonopole (charge--charge) annihilation over monopole-assisted decay of noncontractible pairs (charge--dipole). This means that (i) the free monopoles in the system vanish more quickly, and, correspondingly, (ii) noncontractible pairs are removed more slowly than in the case of truncated interactions.
Since the plateau forms when there are no free monopoles left in the system, point (i) gives rise to the earlier onset of the plateau, while point (ii) implies that the plateau forms at a higher density.

\item The metastable plateau occurs between (\textit{b})--(\textit{d}) and (\textit{c})--(\textit{d}).
This regime, in which the system contains essentially only noncontractible pairs, spans many orders of magnitude in time at the low final quench temperatures considered in this manuscript.

\item At times after (\textit{d}), noncontractible pairs are able to decay via thermal activation, leading to the demise of the metastable plateau. This occurs at a time $\tau_\text{nc} \sim \exp(\Delta / T)$.
\end{enumerate}
By construction, the decay of the plateau occurs at similar times for the models with truncated and long-range Coulomb interactions between monopoles.
The difference in the decay times between the Coulomb and dipolar models is due to the larger variance in energy barriers for activated decay of the pairs in the latter.
Indeed, one may model the decay of the plateau by assuming a Gaussian distribution of energy barriers, $P(\epsilon)$, with mean $\Delta$ and width $\sigma$. The activated decay of the noncontractible pair density $\eta(t)$ is then approximated as $\eta(t) = \int d\epsilon\,P(\epsilon)e^{-t/\tau(\epsilon)}$, where the decay time $\tau(\epsilon) \propto e^{\epsilon/T}$. The values $\sigma_d \simeq \SI{0.1}{\kelvin}$~\cite{Castelnovo2010}, $\sigma_c \simeq \SI{0.03}{\kelvin}$ and $\sigma_t \simeq 0$, lead to the best fit of the Monte Carlo data (not shown).

Notice that, in systems of finite size, the appearance of a noncontractible plateau in the averaged monopole density is, in fact, unavoidable.
On the one hand, the probability that all free monopoles annihilate before all noncontractible pairs have decayed is finite; and, if this happens, the only decay process left for the noncontractible pairs is activated decay.
On the other hand, even when the last two monopoles in the system are free,
there exists a finite probability of forming a new noncontractible pair, rather than annihilation, when the two monopoles come into nearest-neighbour contact.
The latter process places a hard nonzero lower bound on the density of the noncontractible plateau of $O(1/L^3)$, which is purely a finite size effect.


\begin{figure}[t]
  \centering
  \includegraphics[width=\linewidth]{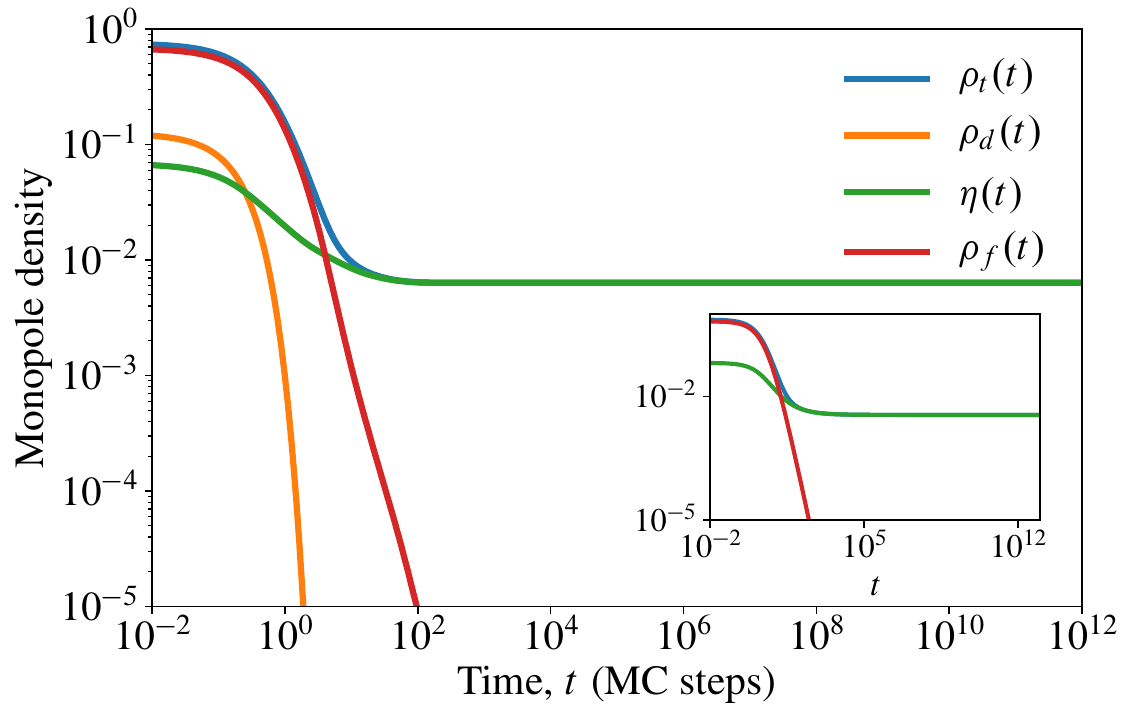}
  \caption{Monte Carlo simulations of charges hopping on the diamond lattice subject to long-range Coulomb interactions [Hamiltonian~\eqref{eqn:charges-hamiltonian-long-range}, system size $L=22$, i.e., $\num{170368}$ spins] from infinite temperature down to zero temperature. Time is expressed in units of Monte Carlo steps per site, and the data are averaged over $\num{4096}$ histories. The analytic solution, \eqref{eqn:nc-density-solution-coulomb}, to the mean field equations for the charge densities is shown in the inset for comparison.}
  \label{fig:randomwalk-coulomb-quench}
\end{figure}

\begin{figure}[t]
  \centering
  \includegraphics[width=\linewidth]{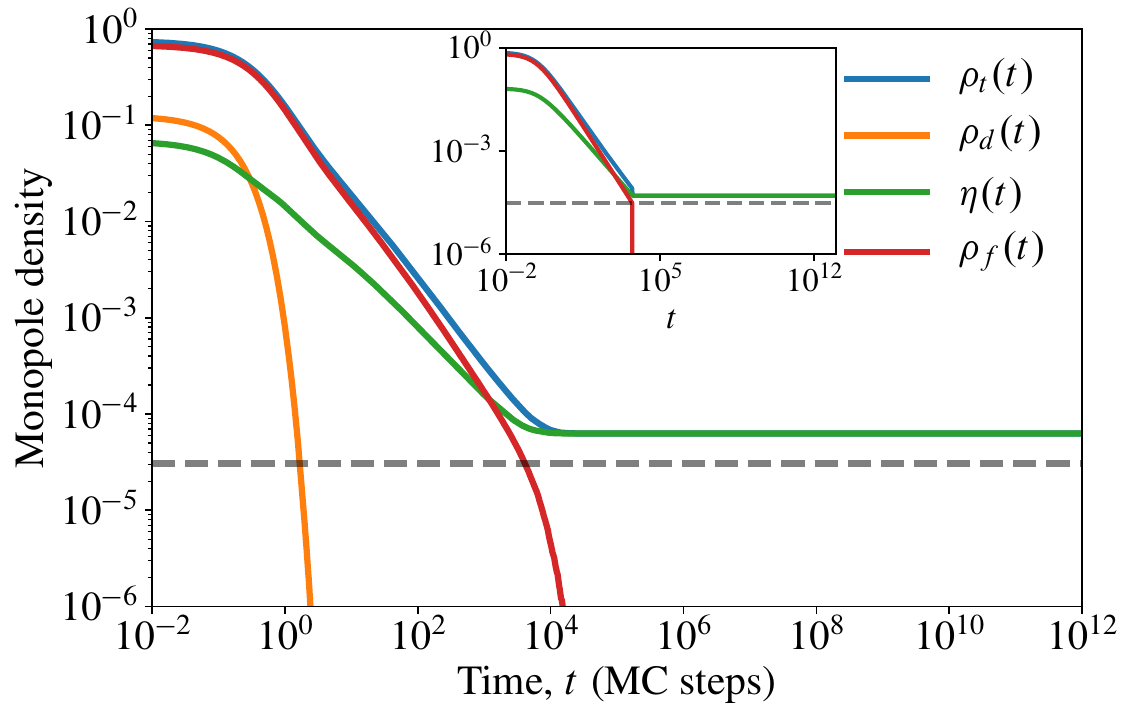}
  \caption{Monte Carlo simulations of charges hopping on the diamond lattice subject to truncated (nearest-neighbour) interactions [Hamiltonian~\eqref{eqn:charges-hamiltonian-truncated}, system size $L=16$, i.e., $\num{65536}$ spins] from infinite temperature down to zero temperature. Time is expressed in units of Monte Carlo steps per site, and the data are averaged over 4096 histories. The analytic solution, \eqref{eqn:nc-density-solution-truncated}, to the mean field equations for the charge densities is shown in the inset for comparison. The dashed lines indicate the threshold density corresponding to the disappearance of free charges in a system of finite size, $\rho_* = 1/N_t$.}
  \label{fig:randomwalk-truncated-quench}
\end{figure}

In order to understand the origin of the plateau and the difference in behaviour between the truncated and long-range interacting models, we ought therefore to look at the finite size scaling behaviour of the plateau density. Figure~\ref{fig: plateau scaling} shows the noncontractible monopole density in the plateau, $\eta_p(L)$, for systems of different sizes (parameterised by the linear system size $L$) and the same final quench temperature $T=\SI{0.06}{\kelvin}$. We perform a fit to the scaling ansatz $\eta_p(L) - \eta_p(\infty) \sim L^{-\nu}$, to extract the exponent $\nu$, the value of the plateau in the thermodynamic limit, $\eta_p(\infty)$, and the constant of proportionality.
The form of this scaling ansatz is justified later in Sec.~\ref{sec:meanfield}, where we show that a power law decay of the free monopole density with time implies power law scaling of the metastable plateau density with system size.
Hence, the scaling ansatz only applies once any transient (non-power-law) behaviour of $\rho_f(t)$ at short times has subsided.
For dipolar interactions between spins, it is not numerically feasible to access system sizes sufficiently large to observe an asymptotic power law decay regime of the free monopole density. We nevertheless provide a fit to the data in this case, but it should be noted that the resulting parameters are subject to some degree of systematic error.
In the case of Coulomb interactions between the monopoles, such asymptotic power law decay of $\rho_f(t)$ is observed in systems of size $L \geq 14$ (i.e., $\num{43904}$ spins), and correspondingly only these data are included in the scaling analysis.

The inset of Fig.~\ref{fig: plateau scaling} demonstrates that the metastable plateau in the truncated case is indeed a finite size effect: The number of noncontractible pairs in the plateau increases subextensively with the size of the system, $\nu \simeq 2.46$, and the density $\eta_p(\infty)$ is consistent with a vanishing value in the thermodynamic limit.
By contrast, in the case of long-range interactions, the number of noncontractible pairs in the plateau scales \emph{extensively} with system size, with subleading, subextensive contributions. Hence, the density of the plateau in the long-range case tends asymptotically towards a finite value, also shown in Fig.~\ref{fig: plateau scaling}. The subextensive corrections give rise to the $L$-dependence of the plateau density. The finite size scaling exponent in this case is $\nu = 0.9(3)$.

We shall summarise these results and attempt to understand the origin of the different behaviours and exponents by modelling the time evolution of the system using mean field population dynamics in Sec.~\ref{sec:meanfield}.
%
%

\subsection{Charges on diamond lattice}
\label{sec:MC:charges}


\subsubsection{Long-range Coulomb interactions}
\label{sec:charges:long-range-coulomb}

Moving to the charge description, characterised by the long-range charge model Hamiltonian~\eqref{eqn:charges-hamiltonian-long-range}, $H^{\rm CM}_c(\{Q_a\})$, we obtain the results shown in Fig.~\ref{fig:randomwalk-coulomb-quench} for a thermal quench down to zero temperature.
As long as the final quench temperature satisfies $T \lesssim E_\text{nn}/L^2$, the dominant effect of changing temperature is to modify the long-time activated decay of the plateau. We therefore focus on the limit of zero temperature for simplicity.

The initial distribution of the charges is set using an infinite temperature distribution of spins on the bonds of the diamond lattice, i.e., using the same initial conditions as in Sec.~\ref{sec:numerics:CSI}.
After initialisation of the system, all reference to an underlying spin configuration is removed, and the time evolution is determined by the dynamical rules laid out in Sec.~\ref{sec:models:charges}.
The most significant difference therefore between the charge model and spin ice systems is the blocked directions imposed by the spins in the latter.
As in the case of the spinful simulations, we measure the various monopole densities as functions of time after the thermal quench and average over histories.

In this case, we observe a plateau that occurs at finite density and which persists indefinitely since the noncontractible pairs cannot undergo activated decay at zero temperature. However, contrasting Figs.~\ref{fig: Coulomb vs dipolar} and~\ref{fig:randomwalk-coulomb-quench}, there are some quantitative differences between the dynamics of the charge and the spin models. In particular, the decay of free monopoles occurs much more quickly in the charge model given the same type of interactions.
This implies that the onset of the plateau occurs significantly earlier in time than the corresponding model in CSI (cf. Fig.~\ref{fig: Coulomb vs dipolar}).


\subsubsection{Truncated interactions}

As shown in Fig.~\ref{fig:randomwalk-truncated-quench}, in the case of truncated interactions between charges, Eq.~\eqref{eqn:charges-hamiltonian-truncated}, we again observe a plateau that occurs at later times and at lower densities than in the long-range interacting charge model (Fig.~\ref{fig:randomwalk-coulomb-quench}).
The free charge density decays approximately as $1/t$ in the long-time limit, i.e., after the double charges have been removed from the system, while the noncontractible pair density also decays as a power law in time, but with a smaller exponent.
The power law decay of these quantities is cut off when the free monopoles drop below $O(1/L^3)$ density, as indicated by the dashed line in Fig.~\ref{fig:randomwalk-truncated-quench}.
The noncontractible pairs that remain in the system can only further decay by thermal activation and the noncontractible plateau is thus established when the free monopole density crosses this threshold.


\subsubsection{Comparison and finite size scaling}

\begin{figure}
  \centering
  \includegraphics[width=\linewidth]{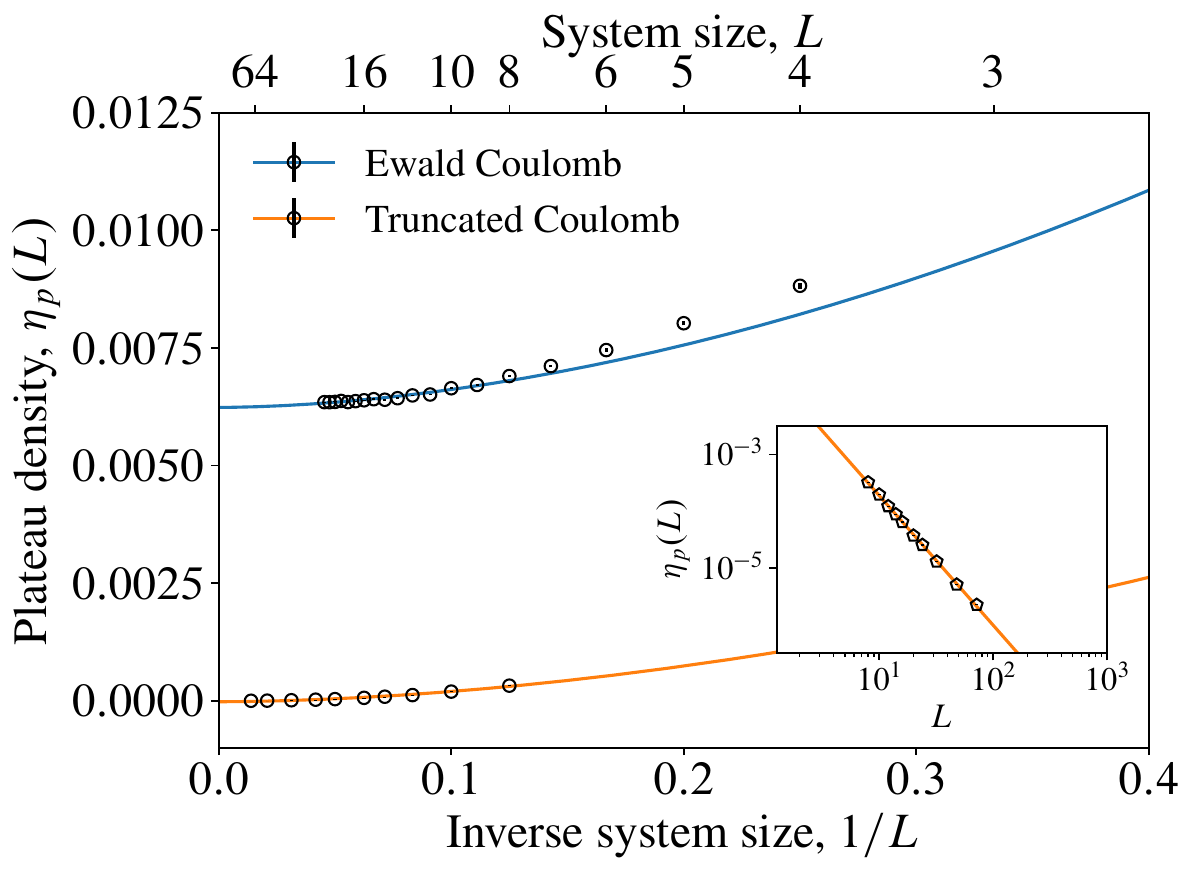}
  \caption{Finite size scaling of the noncontractible plateau density $\eta_p(L)$ for the case of charges hopping on the diamond lattice subject to long-range Coulomb and truncated interactions.
  The data are averaged over at least 4096 histories.
  The lines are fits to the scaling ansatz $\eta_p(L) - \eta_p(\infty) \sim L^{-\nu}$, while the symbols represent the Monte Carlo data.
  The corresponding error bars are smaller than the width of the fit lines.
  As in CSI, the case of truncated interactions ($L=6$--$72$ inclusive) is consistent with a vanishing plateau density in the thermodynamic limit, verified by the log--log plot of plateau density against linear system size in the inset. Conversely, the long-range Coulomb case ($L=4$--$22$ inclusive) exhibits a nonvanishing plateau density in the thermodynamic limit: $\eta_p(\infty)=6.24(2)\times 10^{-3}$.}
  \label{fig:nc-plateau-scaling-charges}
\end{figure}

The finite size scaling of the plateau in the case of charges hopping on the diamond lattice, contrasting the behaviour of Eqs.~\eqref{eqn:charges-hamiltonian-long-range} and~\eqref{eqn:charges-hamiltonian-truncated}, is presented in Fig.~\ref{fig:nc-plateau-scaling-charges}. We again observe that the long-range interacting case tends towards a finite plateau density in the thermodynamic limit, while the plateau is merely a finite size effect in the case of truncated interactions between the charges, i.e., $\lim_{L \to \infty}\eta_p(L) = 0$ with $\nu \simeq 2.28$.

These findings corroborate the conclusions of Sec.~\ref{sec:CSI:finite-size-scaling} pertaining to classical spin ice. In particular, that the plateau is not a finite size effect in the case of long-range Coulomb interactions between charges. Since the subleading corrections decay more quickly in the charge description, $\nu = 1.8(4)$, we are able to make this claim on even stronger terms.

The fact that the finite size scaling of the plateau, i.e., the exponent $\nu$, differs significantly between the spinful and charge descriptions for the long-range case, while it is very similar between the spinful and charge descriptions for truncated interactions, is a puzzle that we shall attempt to understand in Sec.~\ref{sec:meanfield}.
Indeed, we will see that one can achieve a great deal of analytical insight into the observed behaviour by means of appropriate mean field modelling.
%
%


\section{\label{sec:summary}\label{sec:meanfield}
Summary and mean field modelling}

From our simulations we see that the behaviour of the four models in question is visibly similar. The key differences are:
(i) the finite size scaling of the plateau is consistent with a finite versus a vanishing value in the thermodynamic limit in the case of long-range versus truncated interactions, respectively, both in CSI and the CM; moreover, in the case of long-range interactions,
(ii) the decay of $\rho_f(t)$ is notably faster, and the variation with system size $L$ is stronger (i.e., $\nu$ is significantly larger), in the charge simulations than in the spin ice simulations.

\begin{figure}
  \centering
  \includegraphics[width=0.5\linewidth, valign=c]{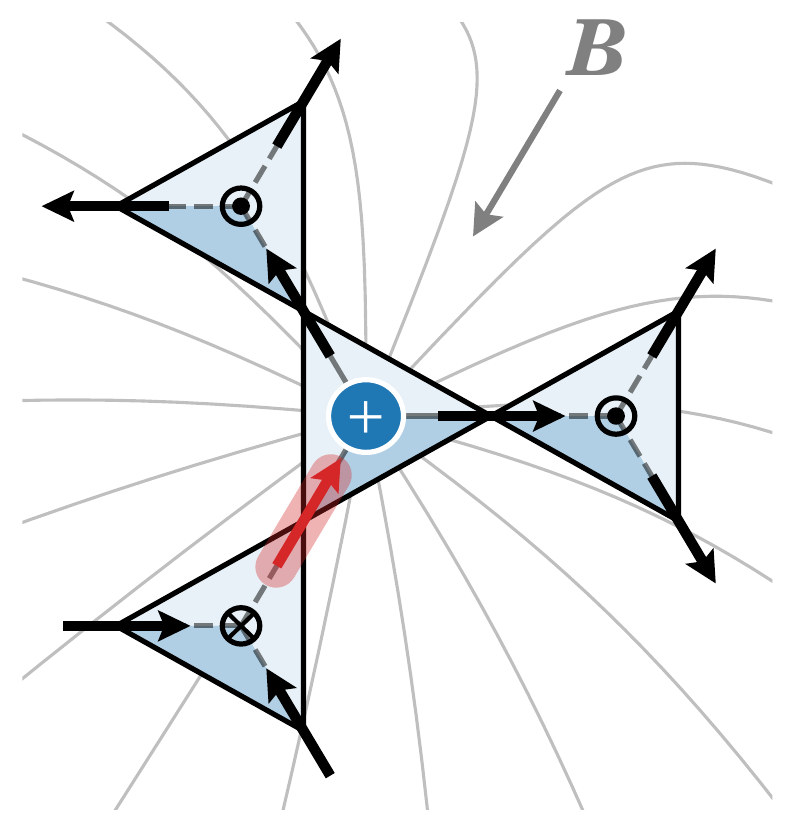}
  \caption{Schematic depiction of a blocked direction for a free monopole.
  The isolated monopole is unable to move along the bond containing the minority spin (shown in red), since its reversal would lead to the creation of a double charge on the central tetrahedron.
  Blocked directions have a significant quantitative impact on the dynamics of monopoles subject to long-range interactions by instantaneously pinning some finite fraction of free monopoles.
  The local magnetic field $\v{B}$ determines which direction(s) lower the energy of the system; if this direction is unique and coincides with the blocked direction (as in the figure), then the monopole is pinned and cannot move along any direction without overcoming a finite energy barrier.}
  \label{fig:blocked-direction-tetra}
\end{figure}

Regarding the discrepancy in the decay of the free monopole density, highlighted in point (ii) above, the most significant difference between the dynamics of the two models in the regime where monopoles are sparse is the existence of blocked directions in classical spin ice~\footnote{One may also wonder whether the differences in the short-time dynamics affect significantly the asymptotic decay of the free monopole density. This has been ruled out by changing between the dynamics generated by long-range classical spin ice and the long-range charge model at some later time, say $t=10\,$MC steps (data not shown).}. That is, for a given (isolated) free monopole, there always exists one of four directions (corresponding to the minority spin) along which the monopole cannot hop, as shown schematically in Fig.~\ref{fig:blocked-direction-tetra}. Assuming that the direction of the local Coulomb field is distributed randomly over the unit sphere, the fraction of charges which are unable to lower their energy due to blocking is $\Omega_b/4\pi$, where $\Omega_b$ is the solid angle for which there is a positive projection onto exactly one of the local basis vectors $\{ \v{e}_i \}$. This leads to a probability
\begin{equation}
  p_b = \frac{\Omega_b}{4\pi}  = \frac{3}{2\pi}\left[ \frac{\pi}{3} - \arctan\sqrt{2} \right] \simeq 4.4\%
  \, ,
  \label{eqn:blocking-fraction}
\end{equation}
for a given free monopole to be pinned (at zero temperature) due to blocking, as shown in Appendix~\ref{app: blocked directions}.
In addition, even when the monopole is not pinned, the available phase space for motion is reduced by blocking.
Notice that~\eqref{eqn:blocking-fraction} underestimates the effect of pinning, because at the lattice scale the direction of the Coulomb interaction is correlated with the bond directions, which violates the assumption of uniformity over the unit sphere. Hence, we conclude that a finite fraction of monopoles, lower-bounded by~\eqref{eqn:blocking-fraction}, are instantaneously~\footnote{Since the spatial distribution of monopoles changes with each Monte Carlo step, the effect of pinning is transient---those monopoles which are pinned at one time may later become unpinned depending on the distribution of free monopoles.
Estimating the relevant timescale is generally difficult,
but one may expect that changing the angle of the local force acting
on a given monopole requires a rearrangement of
the spatial distribution of free monopoles on the order of their
typical separation, which takes a characteristic time
$\sim \rho^{-1/3}$.} pinned in the spinful description due to the interplay of interactions and blocked directions. It is then reasonable to expect that the free monopole density decays more slowly in the presence of such pinned charges.
While this is an interesting aspect of stochastic processes in spin ice that warrants further investigation (maybe by including some effective disorder in the relevant equations governing the dynamics of the charges), it is beyond the scope of the present paper.
We shall nonetheless see below that this effect plays a key quantitative role in the difference between long-range CSI and CM results.

The scaling fits to the Monte Carlo data $\eta_p(L) - \eta_p(\infty) \sim L^{-\nu}$ give the values summarised in Tab.~\ref{tab:finite-size-scaling}. In the following, we show how one can understand this behaviour qualitatively and sometimes even quantitatively using mean field population dynamics of reaction diffusion processes.
\begin{table}[b]
\begin{tabularx}{\linewidth}{@{}p{1.5cm}p{2cm}XX@{}}
\dtoprule
Model                & Interactions & \begin{tabular}[c]{@{}l@{}}Plateau value,\\ $\eta_p(\infty)$\end{tabular} & \begin{tabular}[c]{@{}l@{}}Scaling exponent,\\ $\nu$\end{tabular} \\ \midrule
\multirow{2}{*}{CSI} & truncated    & 0                         & $2.46(1)$          \\
                     & long-range   & $4.7(9) \times 10^{-3}$          & $0.9(3)$         \\ \cmidrule(l){1-4}
\multirow{2}{*}{CM}  & truncated    & 0                         & $2.28(2)$          \\
                     & long-range   & $6.24(2) \times 10^{-3}$          & $1.8(4)$         \\ \dbottomrule
\end{tabularx}
\caption{Summary of finite size scaling results for both systems and both types of interaction between the tetrahedral charges. The scaling ansatz $\eta_p(L) - \eta_p(\infty) \sim L^{-\nu}$ was used to obtain the values shown in the table.}
\label{tab:finite-size-scaling}
\end{table}
This allows us to model the time evolution of the monopole/charge densities and to obtain estimates of the finite size scaling exponents to compare with our numerical results.
%
%

\subsection{\label{sec:meanfield:short time}
Short-time dynamics
           }
If we want to describe the simulations in terms of reaction-diffusion processes between (effective) particles, we ought to consider in principle five different species: positive and negative single and double charges, and noncontractible pairs.
The noncontractible pairs are immobile, pinned to the bond on which they form, and can undergo the activated and monopole-assisted decay processes discussed previously.
Single charges are able to move freely throughout the lattice (neglecting the effects of spin blocking/pinning).
The double charges can either decay spontaneously into two single charges of the same sign, if adjacent to an empty site, or they can be hit by a single charge of the opposite sign and decrease their charge by one, thus producing a single (mobile) charge.
Finally, two adjacent double charges of opposite sign can decay to form a noncontractible pair by flipping the intervening spin.
All decay processes involving double charges reduce the energy of the system, and thence are able to occur spontaneously, even at zero temperature.

The rate of decay of double monopoles does depend on the free monopole density; however it is easy to convince oneself that the `phase space' for decay (either spontaneous or monopole-assisted) is always larger than that for processes which preserve the number of double charges, and it becomes progressively more so as the free monopoles decay in time.
Their evolution thus rapidly decouples from the other species and becomes exponentially fast in time: asymptotically $\rho_d(t) \propto e^{-7t/2}$, as argued in Appendix~\ref{app: double charges}, which appears to fit well all simulations.

The single charges that are produced in the decay of double charges merely become a known time-dependent source term in the corresponding equation governing their density; as we see from the simulations, this contribution becomes irrelevantly small for $t \gtrsim 1$. When looking at the total or free monopole/charge densities, the double charges contribute towards the `hump' observed at short times, before the onset of the asymptotic power-law behaviour.
In Appendix~\ref{app: double charges} we discuss this in greater detail, and we show explicitly that the double charge contribution indeed does not affect the asymptotic scaling behaviour we are interested in understanding, affecting only the density of the noncontractible plateau.

For these reasons, in the following, we shall ignore the double charges altogether and focus on the three remaining species of particle: positively and negatively charged free monopoles living on the sites of a diamond lattice, with densities $\rho_q(t)$ (charge $q=\pm$); and immobile noncontractible pairs living on the bonds, with density $\eta(t)$. The equations determining their dynamics are presented and analysed in the following sections.
%
%

\subsection{\label{sec:meanfield:truncated}
Truncated interactions
           }
The mean field equations (i.e., neglecting spatial fluctuations) describing the time evolution of the monopole densities in the case of truncated interactions between monopoles are (for $q=\pm$)
\begin{align}
  \dv{\rho_q}{t} &= -\mc{K} \rho_+ \rho_- \, , \label{eqn:mean-field-modelling-1} \\
  \dv{\eta}{t} &= -\frac{\mc{R}}{2}(\rho_+ + \rho_-) \eta + \mc{K}'\rho_+\rho_-
	\, .
  \label{eqn:mean-field-modelling-2}
\end{align}
Equation~\eqref{eqn:mean-field-modelling-1} describes the annihilation of oppositely charged free monopoles, which occurs with rate $\mc{K}$. The first term in Eq.~\eqref{eqn:mean-field-modelling-2}, with rate $\mc{R}$, describes the monopole-assisted decay of noncontractible pairs---a free monopole annihilates the member of a noncontractible pair with the opposite sign. Such a process removes two monopoles previously forming a noncontractible pair, but preserves the number of free monopoles in the system, and therefore does not appear in~\eqref{eqn:mean-field-modelling-1}. Finally, the second term in~\eqref{eqn:mean-field-modelling-2} describes the probabilistic formation of noncontractible pairs when two oppositely charged monopoles come into nearest-neighbour contact.
As we want to understand the origin and scaling behaviour of the noncontractible pair plateau, we are not interested in the very long-time behaviour of the system. We have therefore disregarded the terms corresponding to the activated decay of the noncontractible pairs.
Equivalently,~\eqref{eqn:mean-field-modelling-1} and~\eqref{eqn:mean-field-modelling-2} describe the zero-temperature dynamics of the system.

Charge neutrality ensures that $\rho_+(t) = \rho_-(t)$ for all times, allowing us to solve \eqref{eqn:mean-field-modelling-1} for the time evolution of the free monopole densities $\rho_q(t)$:
\begin{equation}
  \rho_q(t) = \frac{\rho_{q}^0}{1 + \mc{K}\rho_q^0 t}
  \, ,
\end{equation}
where $\rho_q^0 \equiv \rho_q(0)$.
This solution may then be substituted into~\eqref{eqn:mean-field-modelling-2} describing the noncontractible monopole density $\eta(t)$
\begin{equation}
  \dv{\eta}{t} + \mc{R}\rho_q(t) \eta = \mc{K}'\rho_q^2(t)
  \, ,
\end{equation}
which can also be solved exactly to give
\begin{eqnarray}
  \eta(t) &=&
  \frac{(\mc{K}'/\mc{K}) \rho_q^0}{(\mc{R}/\mc{K} - 1)(1+\mc{K}\rho_q^0 t)}
  \nonumber \\
  &+&
  \left[ \eta_0 - \frac{(\mc{K}'/\mc{K}) \rho_q^0}{\mc{R}/\mc{K} - 1} \right]
  \frac{1}{(1+\mc{K}\rho_q^0 t)^{\mc{R}/\mc{K}}}
  \, .
  \label{eqn:nc-density-solution-truncated}
\end{eqnarray}

Evidently, the long-time behaviour of the noncontractible monopole density $\eta(t)$ depends crucially on the ratio of rate constants $\mc{R}/\mc{K}$. If $\mc{R}/\mc{K} < 1$, then the second term in~\eqref{eqn:nc-density-solution-truncated} dominates at long times and the noncontractible pairs decay more slowly than the free monopoles, as is observed in the numerics, illustrated in particular in Figs.~\ref{fig: truncated Coulomb} and~\ref{fig:randomwalk-truncated-quench} (this is also consistent with the analytic estimates of $\mc{R}/\mc{K}$ that we present below).

In the thermodynamic limit, these equations predict that there is no plateau in the noncontractible pair density since both $\rho_q(t)$ and $\eta(t)$ may become arbitrarily small. However, in a system of finite size containing $8 L^3$ tetrahedra, the decay of $\rho_q(t)$ is cut off when the free monopole density reaches $O(1/L^3)$: $\rho_q(t_*) \sim L^{-3}$, i.e., at a time $t_* \sim L^3$ corresponding to the removal of \emph{all} free monopoles in a finite system. If the noncontractible pair density decays more slowly, as is the case for $\mc{R}/\mc{K} < 1$, there is still a finite density of noncontractible pairs present in the system at $t_*$, and they can further decay only via thermal activation. The value of this density scales as $\eta(t_*) \sim t_*^{-\mc{R}/\mc{K}}$ for sufficiently large $t_* \gg (\mc{K}\rho_q^0)^{-1}$, allowing us to deduce the leading order term in the dependence of the noncontractible plateau on system size:
\begin{equation}
  \eta(t_*) \sim L^{-3\mc{R}/\mc{K}} \, ,
\end{equation}
and therefore extract the exponent $\nu = 3\mc{R}/\mc{K}$.

We can estimate the ratio $\mc{R}/\mc{K}$ from the microscopic details of our system as the product of two contributions,
\begin{equation}
  \frac{\mc{R}}{2\mc{K}} = \frac{N_\mc{R}}{N_\mc{K}} \cdot \frac{\tau_\mc{K}}{\tau_\mc{R}} \simeq \frac{3}{4} \cdot \frac{1}{2} \, .
  \label{eqn:scaling-exponent}
\end{equation}
The first factor in~\eqref{eqn:scaling-exponent}, $N_\mc{R} / N_\mc{K}$, comes from the fact that a free monopole has 4 adjacent free legs along which another free monopole may approach, while a noncontractible pair has only 3 (one of the four total legs being blocked by the other member of the pair)~\footnote{We note that in the spinful description, the rate constant $\mc{K}$ includes the formation of noncontractible pairs in addition to annihilation events, and so blocked directions do not alter this argument to leading order.}. Therefore the factor $3/4$ encodes the relative sizes of the basins of attraction in the two cases.
The second factor $\tau_\mc{K} / \tau_\mc{R}$ derives from the ratio of timescales---in the case where two free monopoles are approaching one another, both are mobile, while in the case of a free monopole approaching a noncontractible pair, the noncontractible pair is pinned and only the free monopole is mobile. This leads to a factor of 2 difference in the (random walk) timescales for the two processes.
The factor of $1/2$ on the left hand side of~\eqref{eqn:scaling-exponent} originates from the definition of $\mc{R}$ in~\eqref{eqn:mean-field-modelling-2}.
We therefore estimate that $\mc{R}/\mc{K} \simeq 3/4$, and correspondingly the noncontractible plateau scales approximately as
\begin{equation}
  \eta(t_*) = \eta_p(L) \sim L^{-9/4} \, ,
\end{equation}
in the case of truncated interactions between charges.

This estimate can be improved upon by examining larger clusters.
Indeed, including next-nearest neighbours in the cluster, the presence of blocked directions leads to a small correction to the finite size scaling exponent in the case of CSI, as shown in Appendix~\ref{app:finite-size-exponents}, while it remains unchanged for the CM:
\begin{equation}
  \nu_{\text{CSI}} = \frac{90}{37} \simeq 2.43 \, , \quad \nu_{\text{CM}} = \frac{9}{4} = 2.25
  \, .
\end{equation}
These exponents are consistent with the values $\nu=2.46(1)$ and $\nu=2.28(2)$ obtained from the Monte Carlo data in Figs.~\ref{fig: plateau scaling} and~\ref{fig:nc-plateau-scaling-charges}, respectively. Note that the absolute values of $\mc{R}$ and $\mc{K}$ differ substantially between CSI and the CM due to the presence of blocked directions in the former, but their \emph{ratio} remains essentially the same.

We are now able to understand why the spinful and charge descriptions exhibit quantitatively similar behaviour. In both cases, the charges exhibit diffusive motion (until they become nearest neighbours, at which point they deterministically annihilate). The numerical results suggest that the annealed (random) blocked directions do not significantly affect the diffusive motion of the charges, and therefore do not alter the form of the decay of the free monopole density.
This is because the motion of monopoles across the system (i.e., beyond nearest-neighbour separation) is not subject to any energetic bias controlling the direction of their motion. Hence, the insertion of blocked directions at random has little effect on the purely random motion of charges when averaged over histories---no monopoles are instantaneously pinned due to blocking.
This is also evidence of the fact that entropic interactions in CSI due to the underlying spins do not play a significant role in the evolution of the monopole density following a thermal quench.
The free monopole density decays as $1/t$ in both CSI and the CM with truncated interactions, and we consequently obtain a vanishing plateau in the thermodynamic limit. Further, the value of $\nu$ is set by the ratio of the rates of monopole-assisted decay to free monopole annihilation, which is common to both descriptions, up to small corrections which result from the impact of blocked directions on the microscopic annihilation process.


\subsection{Long-range Coulomb interactions}

In Sec.~\ref{sec:meanfield:truncated} we were able to develop a rather complete understanding of the case of truncated interactions, which largely hinged on the $1/t$ scaling of the free monopole density.
We would now like to study how the behaviour changes in the presence of long-range interactions.
One could naively try to introduce them at the level of the reaction diffusion equations; however,
this is known to recover the law of formal kinetics at long times, i.e., $1/t$ behaviour of $\rho_f(t)$, which leads to the same conclusion of a vanishing plateau value in the thermodynamic limit.
This is however in contradiction with the observation that $\rho_{f}(t)$ decays faster than $1/t$ in our Monte Carlo simulations of long-range interacting systems (and with the observation of a finite value for the metastable plateau).

As is often the case, the devil lies in the details. In order to observe a long-lived metastable plateau, we need to quench to very low temperatures, $T \ll J_\text{eff}$. In a discrete system with long-range interactions and finite lattice spacing, the hydrodynamic description of Refs.~\onlinecite{Ovchinnikov2000}
and~\onlinecite{Ginzburg1997} does not always apply to the Monte Carlo time evolution of our simulations.
Take for example the limiting case of a quench to zero temperature. The quasiparticles move only downwards or across in energy, $\delta E \leq 0$, and they move at `terminal velocity' (i.e., one lattice spacing per unit time) irrespective of the strength of the force acting upon them. On the contrary, the hydrodynamic description applies when the Monte Carlo process is a (lightly) biased random walk, $|\delta E| \ll T$, and the equations of motion approximately take the familiar overdamped form where the velocity of the particles is proportional to the force acting on them. This is how our simulations violate the law of formal kinetics (at intermediate times) and achieve a decay of free monopole density which is faster than $1/t$ at the low temperatures studied in this manuscript.

Modelling the strictly-biased motion at terminal velocity is a tall order.
However, at mean field level, one can put forward the following approximate argument: the free monopole density decays with a time constant given by the time taken to travel at terminal velocity to the next free monopole, some characteristic distance $\rho^{-1/d}$ away, namely $\tau_{\rm tv} \sim \rho^{-1/d}$, where $d$ is the dimensionality of the system. Then we have
\begin{equation}
\dv{\rho}{t} \propto - \frac{\rho}{\tau_{\rm tv}}
\qquad
\Rightarrow
\qquad
\rho(t) \sim 1/t^d
\, .
\label{eq: terminal velocity decay}
\end{equation}
This behaviour is in very good agreement with the $\rho_{f}(t)$ decay observed in the CM with long-range interactions if one neglects the formation of noncontractible pairs. We shall delay the discussion of the CSI case to later in this section.

In the absence of long-range interactions, there are no forces beyond a fixed finite separation between monopoles and they perform an unbiased random walk, even at zero temperature.
It then takes a characteristic time, $\rho^{-1}$,
corresponding to the time taken for a monopole to explore its characteristic volume in three dimensions, to come in contact and annihilate with another monopole. In this case, $\tau_{\rm tv}$ should be replaced by $\tau_{\rm rw} \sim \rho^{-1}$ and one recovers the $1/t$ scaling obtained more rigorously in Sec.~\ref{sec:meanfield:truncated}.

In order to express all these considerations more formally, and to take into account explicitly the noncontractible pair density $\eta(t)$, which has been ignored thus far, it is convenient to introduce the following phenomenological reaction diffusion equations
\begin{align}
  \dv{\rho_q}{t} &= -\mc{K} [\rho_+(t) \rho_-(t)]^{(1+\beta)/2}   \, , \label{eqn:mean-field-coulomb-1} \\
  \dv{\eta}{t} &= -\frac{\mc{R}}{2}(\rho_+ + \rho_-) \eta - \frac{\mc{K}'}{\mc{K}} \dv{\rho_q}{t}
	\, ,
	\label{eqn:mean-field-coulomb-2}
\end{align}
with the parameter $\beta \leq 1$ (with $\beta = 1$ corresponding to the truncated case, and $\beta = 1/3$ corresponding to the terminal velocity argument given above, neglecting the effect of nonzero $\eta$)~\footnote{We have defined $\beta$ in this way in order to make the formulae that follow neater and more compact.}.

Using charge neutrality $\rho_+(t) = \rho_-(t)$, the first of these equations gives rise to a free monopole density
\begin{equation}
  \rho_q(t) = \frac{\rho_q^0}{(1 + \beta\mc{K}_0{\rho_q^0}t)^{1/\beta}}
  \, ,
\end{equation}
where we have defined for convenience $\mc{K}_0 \equiv \mc{K}(\rho_q^0)^{\beta-1}$.
The parameter $\beta$ sets the asymptotic rate of decay of the free monopole density in the system: $\rho_q(t) \sim t^{-1/\beta}$. This decay is faster than the truncated case ($\rho_q \sim 1/t$) when $\beta < 1$.
Defining
\begin{align}
  \Theta(t) &= \int_0^t \mathrm{d}t^\prime \, \rho_q(t^\prime)  \\
  &= \frac{1}{1-\beta}\frac{1}{\mc{K}_0} \left[ 1 - (1+\beta\mc{K}_0\rho_q^0 t)^{(\beta-1)/\beta} \right] \label{eqn:theta-coulomb}
  \, ,
\end{align}
the solution for the noncontractible monopole density may be written as
\begin{equation}
  \eta(t) = e^{-\mc{R}\Theta(t) } \left[ \eta_0 + \int_0^{t} \mathrm{d}t^\prime e^{\mc{R}\Theta(t')} \mc{K}^\prime [\rho_+(t') \rho_-(t')]^{(1+\beta)/2}  \right]
  \, .
  \label{eqn:coulomb-nc-pair-solution}
\end{equation}
It is possible to obtain an analytic expression for $\eta(t)$ by expressing the integral in~\eqref{eqn:coulomb-nc-pair-solution} in terms of the incomplete Gamma function, which is presented in Appendix~\ref{sec:meanfield-solution}. Since, for $\beta < 1$, $\Theta(t)$ tends towards a constant at large times, the solution for $\eta(t)$ exhibits a plateau at finite density, $\eta(t) \to \eta_\infty$, as $t \to \infty$. The density at which this plateau occurs is
\begin{equation}
  \eta_\infty = e^{- \alpha\mc{R}/\mc{K}_0} \left\{ \eta_0 + \rho_q^0 \frac{\alpha\mc{K}'}{\mc{K}} e^{\alpha\mc{R}/\mc{K}_0} \left[ \frac{\alpha\mc{R}}{\mc{K}_0} \right]^{-\alpha} \gamma\left( \alpha,  \frac{\alpha\mc{R}}{\mc{K}_0} \right) \right\}
  \, ,
  \label{eqn:thermodynamic-plateau}
\end{equation}
where $\alpha \equiv 1/(1-\beta)$, and $\gamma(s, x)$ is the lower incomplete gamma function. Hence, the value of the plateau is exponentially sensitive to the ratio of rate constants $\mc{R}/\mc{K}_0$, and vanishes as $\beta \to 1^-$ (i.e., $\alpha \to \infty$).

At sufficiently large times,
\begin{equation}
  \eta(t) \simeq \eta_\infty \left[ 1 + \frac{\alpha\mc{R}}{\mc{K}_0} (\beta\mc{K}_0\rho_q^0 t)^{(\beta-1)/\beta} \right]
  \, .
\end{equation}
The finite size scaling of the noncontractible plateau then follows from the fact that the free monopole decay is cut off at a time $t_*$, defined by $\rho_q(t_*) \sim L^{-3}$. As before, $t_*$ equals the time at which free monopoles are completely removed from a system of finite size. This gives $t_* \sim L^{3\beta}$ and correspondingly the finite size scaling of the plateau satisfies
\begin{equation}
  \eta(t_*) - \eta_\infty \sim t_*^{-(1-\beta)/\beta} \sim L^{-3(1-\beta)}
  \, .
\end{equation}
The scaling exponent of the plateau, $\nu$, can therefore be directly related to the exponent $\beta$ which quantifies the asymptotic rate of decay of the free monopole density,
\begin{equation}
    \nu = 3(1-\beta)
    \, .
\label{eq: scaling nu beta}
\end{equation}
This relationship is consistent with the discrepancy between the finite size scaling exponents in the long-range interacting CSI and CM cases: The rapid decay of the free monopole density permitted by the lack of blocked directions in the CM case implies a larger $\beta^{-1}$ and, hence, a larger $\nu$. Indeed, numerically fitting the exponent of the asymptotic free monopole decay, we obtain $\beta^{-1} \simeq 1.4$ and $\beta^{-1} \simeq 2.3$ corresponding, through~\eqref{eq: scaling nu beta}, to scaling exponents $\nu \simeq 0.86$ and $\nu \simeq 1.7$ for the cases of long-range CSI and the CM, respectively. These values are in reasonable agreement with those obtained from the numerical finite size scaling analysis: $\nu = 0.9(3)$ and $\nu = 1.8(4)$.

Notice that the mean field equations~\eqref{eqn:mean-field-coulomb-1} and~\eqref{eqn:mean-field-coulomb-2} can only be expected to hold at asymptotically long times for zero-temperature quenches. For any finite $T$, as the monopoles become sparser, the forces between them become weaker and eventually one reaches the hydrodynamic regime, $|\delta E| \ll T$, discussed earlier, and a $1/t$ decay of $\rho_f(t)$ ensues.
The typical Coulomb interaction felt by a given monopole through the separation $\rho(t)^{-1/d}$ is (in $d=3$ for concreteness)
\begin{equation}
  \ev{E_c(t)} \sim -E_\text{nn} \rho(t)^{1/3}
  \, .
\end{equation}
The corresponding \emph{change} in Coulomb energy when moving a free monopole
to an adjacent site then scales as
\begin{equation}
  \ev{\delta E_c(t)} \sim E_\text{nn} \rho(t)^{2/3}
  \, .
\end{equation}
Assuming $\rho(t) \sim 1/t^{1/\beta}$, the time threshold $\ev{\delta E_c(t)} \sim T$ corresponding to the crossover to $1/t$ decay of $\rho_f$ can then be estimated to scale with temperature as $t_T \sim (E_\text{nn} / T)^{3\beta/2}$~\footnote{%
Taking the terminal velocity limit, $\beta = 1/3$, and using the parameters
for {\DTO}, the thermal crossover occurs at $t_T \sim \SI{0.1}{\second}$.
This allows the system to enter a metastable state dominated by
noncontractible pairs, which then live for a time set by thermal
activation, $e^{\Delta/T} \sim 1\,\text{year}$.
The crossover to the hydrodynamic regime for $t \gg t_T$ occurs only
for sufficiently large system sizes (namely, if the system can
access sufficiently low monopole densities), which correspond to
samples of linear dimension much larger than $\SI{10}{\nano\meter}$}.
The crossover can be observed in our Monte Carlo simulations
at sufficiently high temperatures;
however it is barely visible within the accessible system sizes
and the corresponding plots are not very informative, and we
refrain from showing them here.
From~\eqref{eqn:mean-field-coulomb-2}, we deduce that the noncontractible plateau therefore begins to decay at times $t \gtrsim t_T$.
The rate of decay however vanishes as temperature is lowered, i.e., $\ln\eta \sim - T^{\nu/2} \ln t$.
The zero-temperature limit therefore does not commute with the limit of infinite time. If the latter is taken first, the plateau decays to a vanishing thermodynamic value at large times. If the former is taken first, then a finite plateau survives.
Since the timescale for activated decay of the plateau scales exponentially with temperature, while $t_T$ scales algebraically [at least for a power law decay of $\rho_f(t)$], it will be the case that $t_T < \exp(\Delta/T)$ at the low but nonzero quench temperatures that we considered in this manuscript.
For systems of finite size, the relevant question then becomes whether $t_T$ is larger or smaller than the time $t_*$ that it takes for the free monopole density to become less than $O(1/L^3)$.

We finally note that even at zero temperature the mean field equations will eventually break down at a time corresponding to single charge densities $\rho_q$ at which free charges become so dilute that the bias for free charge--charge annihilation over monopole-assisted decay is removed. We term such a time $t_d$, which may be obtained by comparing $\ev{\delta E_c(t)}$ with the typical energy due to charge--dipole interactions with the noncontractible pairs present in the metastable plateau regime. Once this bias disappears, monopole-assisted decay may once again become favourable and the plateau is able to gradually decay.

The phenomenological model that we have presented illustrates in a simple manner the mechanisms at play, but we note that the precise functional form or even the asymptotic power law decay of the free monopole density implied by the model are not a requirement in order to observe a noncontractible plateau in the thermodynamic limit.
Indeed, at the mean field level, \emph{any} decay of $\rho_f(t)$ faster than $1/t$ will give rise to a plateau in the density of monopoles forming noncontractible pairs.
Even if $\rho_f(t)$ does exhibit a crossover to $1/t$ behaviour at long times, the plateau will still be present in the thermodynamic limit, but will only exist for a finite period of time before it starts to decay.
%
%

\section{\label{sec:conclusions}
Conclusions
        }

Using a combination of Monte Carlo simulations and detailed mean field modelling, we investigated the origin of the metastable plateau that is observed in thermal quenches to low temperatures in classical spin ice~\cite{Castelnovo2010}. Our results show that it is a consequence of the long-range nature of the Coulomb-like interactions between monopoles combined with the system entering a non-hydrodynamic regime which is controlled by nonuniversal lattice physics. The claim that such a plateau may have been observed in recent experiments~\cite{Paulsen2016} therefore provides further compelling evidence for the long-range nature of the interactions between the emergent monopoles in these systems.

In particular, we have shown that when the interactions between the monopoles are truncated to finite range, the plateau reduces to a finite size effect. This is because the free monopoles in the system perform independent random walks (when their density is sufficiently low) leading to a $1/t$ decay of their density with time $t$. Although this is sufficient to create
the ``population inversion'' (in which noncontractible pairs
become the dominant species in the total monopole density),
the slow decay of free monopoles implies that monopole-assisted
decay remains effective and continues to remove noncontractible
pairs from the plateau indefinitely.
On the contrary, in the presence of long-range Coulomb interactions between monopoles, there exists an energetic bias in their motion across the system. At sufficiently low temperatures, which are relevant for the formation of a thermodynamic noncontractible plateau, the system enters a non-hydrodynamic regime in which the monopoles move at terminal velocity in the direction of the local force acting on them.
This combination of long-range interactions and non-hydrodynamic behaviour leads to a rapid decay of the free monopole density, faster than $1/t$ and violating the law of formal kinetics. The decay of free monopoles is then sufficiently rapid to stop the monopole-assisted decay of noncontractible pairs at long times, and therefore one observes a plateau of finite density in the thermodynamic limit.

In this paper we studied the case of sudden quenches to the target temperature. Spin ice systems and materials are well-known to exhibit long relaxation timescales at low temperatures and a relevant and interesting question would be to investigate how much of the phenomenology observed in the present work survives in the case of ramps, where the temperature is lowered continuously to its target value, a question that is indeed of experimental importance. It would be particularly interesting to see if there is a threshold in the ramp speed beyond which the behaviour changes qualitatively. We note however that such studies, which are beyond the scope of the present work, will likely require accessing significantly lower monopole densities and therefore simulating larger system sizes, possibly beyond the current numerical capability.

Our numerical results are in quantitatively good agreement with
analytics from mean field modelling. This may come as a surprise
if one thinks that fluctuations in the charge density ought to
bring about corrections that are not captured by mean field theory.
However, emergent charges in spin ice systems are subject to a
hard-core, hyperuniform constraint in their spatial distribution:
The charges are born out of the underlying spins and one can easily
verify that the maximal net charge that can be accumulated in a
volume $\ell^3$ scales as $\ell^2$ (as opposed to free
charge systems, where the latter can scale as $\ell^3$). As a result,
long-wavelength fluctuations are suppressed, and
one can expect mean field calculations to be in fact rather accurate
in describing spin ice behaviour.
We note that the charge model introduced in this manuscript is
not in general subject to the same constraint. However, we
impose the
same initial conditions as in the spin ice system, which are therefore
hyperuniform. The good agreement with mean field theory
suggests that this seeding is sufficient to maintain hyperuniformity
throughout the time evolution following the quench (at least within
the system sizes and time scales accessible in our simulations).

Given the importance of including exchange interactions between spins
beyond nearest-neighbour separation in describing the equilibrium
(and out-of-equilibrium)
properties of spin ice~\cite{Yavorskii2008, Henelius2016, Borzi2016, Samarakoon2019},
it is pertinent to ask what the effect of such farther-ranged
interactions might be on the thermal quenches
discussed in the present work.
Consider the inclusion of second- and third-neighbour interactions $J_2$ and
$J_3$, respectively. The latter is subdivided into $J_{3a}$ and $J_{3b}$, as described in Appendix~\ref{app:J2-J3}.
In the special case $J_2 = -3J_{3a}$ and $J_{3b}=0$, these interactions
can be summed to give exactly the truncated (nearest-neighbour) interactions
between charges: $\propto J_2 \sum_{\langle ab \rangle} Q_a Q_b$
(in addition to a shift of $J_\text{eff}$, see Appendix~\ref{app:J2-J3}).
The inclusion of such farther-ranged interactions hence
modifies the short-distance
physics of monopoles and leads, for example, to a modification of the barrier to
activated decay of noncontractible pairs.
When the interactions do not satisfy this condition,
we expect nonetheless that the behaviour of the system will remain
qualitatively similar provided that the long-range bias for monopole motion
across the system is active during the transient
(terminal velocity) regime
in which the plateau is established.

Direct observation of the behaviour studied in this work requires experimental probes that measure the monopole density in spin ice materials. One could envisage using the width of the pinch points in the neutron scattering structure factor~\cite{Fennell2009, Henley2010} (with a caveat on the contribution from nearest-neighbour pairs, such as the noncontractible pairs, as discussed in Ref.~\onlinecite{Szabo2019}). Alternatively, small quenches in the magnetic field, and a measurement of the magnetisation $M(t)$ that ensues, give access to the time evolution of the free monopole density, since $dM/dt \propto \rho_f$~\cite{Slobinsky2010}. Further experimental probes of monopole density in spin ice would be very much welcome in this respect.

The potential departure of long-range interacting lattice systems from a hydrodynamic description, and thence from the law of formal kinetics, is somewhat expected: At sufficiently low temperatures, the change in energy incurred by a microscopic discrete update in the system becomes larger than the thermal energy. However, one generally expects this phenomenon to affect only the short-time dynamics, and that at long times the universal hydrodynamic behaviour is recovered. Thermal quenches in spin ice demonstrate that, while this expectation must ultimately be satisfied, the altered nonuniversal, transient dynamics during times $t \lesssim \SI{1}{\second}$ can induce very long-lived metastable states that change the behaviour of the system over a large range of `intermediate' times spanning many orders of magnitude (easily growing to be of the order of $1\,\text{year}$ or longer for experimentally relevant parameters and temperatures).

This phenomenon may play a role in other aspects of the behaviour of spin ice models and materials at low temperature (for example, a departure from hydrodynamic behaviour could be a contributing factor to the deviation from the so-called `quasiparticle kinetics' in Ref.~\onlinecite{Paulsen2016}). It may also be relevant to other long-range interacting natural and artificial lattice systems of interest.
%
%

\begin{acknowledgments}
CC is particularly grateful to R.~Moessner, with whom the seed ideas behind this project were formulated. The authors would also like to thank G.~Goldstein, P.~Krapivsky, and C.~Laumann for insightful discussions.
This work was supported in part by the Engineering and Physical Sciences Research Council (EPSRC) Grants No.~EP/K028960/1,~EP/M007065/1, and~EP/P034616/1.
This project was carried out using resources provided by the Cambridge Service for Data Driven Discovery (CSD3) operated by the University of Cambridge Research Computing Service (\href{http://www.csd3.cam.ac.uk/}{\color{black}http://www.csd3.cam.ac.uk/}), provided by Dell EMC and Intel using Tier-2 funding from the Engineering and Physical Sciences Research Council (capital grant EP/P020259/1), and DiRAC funding from the Science and Technology Facilities Council (\href{www.dirac.ac.uk}{\color{black}www.dirac.ac.uk}).
\end{acknowledgments}
%
%
\appendix

\section{\label{app:simulation-details}
Simulation Details
        }
Ewald summation leads to the following expression for the Coulomb energy of a set of interacting charges $\{ q_a \}$ and their periodic images
\begin{equation}
    E_c(\{ q_a \}) = \sum_{a < b} q_a K_{ab} q_b + \mu \sum_a q_a^2
    \, ,
\end{equation}
where we have defined $K_{aa} \equiv 0$, $\forall a$, having separated out the diagonal terms, which may be absorbed into the effective chemical potential for charges. Supposing that we flip a spin $S_i$, the charges on the two adjacent tetrahedra, labelled by $a$, $b$, are modified: $q_a \to Q_a$, and $q_b \to Q_b$. The change in Coulomb energy when flipping this spin is therefore
\begin{eqnarray}
\delta E_{c}(Q_a, Q_b) &=&
  \sum_{c\, : \, q_c \neq 0}
	  \left[ \delta q_a K_{a c} + \delta q_b  K_{b c}  \right] q_c
\nonumber \\
&+& \delta q_a K_{ab} \delta q_b +
    \mu \left[ \delta(q_a^2) + \delta(q_b^2) \right]
\, ,
\label{eqn:coulomb-spin-flip}
\end{eqnarray}
where $\delta q_a = Q_a - q_a$, and $\delta(q_a^2) = Q_a^2 - q_a^2$. Such an expression already represents an improvement over the conventional dipolar Monte Carlo code---one needs only to sum over the nonzero charges, which are dilute in the metastable plateau.

However, one can further speed up the computation of the Coulomb energy by considering the \emph{change} in the Coulomb spin flip energies when going from time step $t \to t+1$.
Suppose that in the Waiting Time Monte Carlo (WTMC) update at time $t$, spin $S_j$, adjacent to tetrahedra $c$,~$d$, was flipped. We then propose flipping $S_i$, adjacent to tetrahedra $a$,~$b$. If there is no overlap between tetrahedra $a$,~$b$ and $c$,~$d$ (i.e., none of $a$, $b$, $c$, $d$ are equal), the change in spin flip energy between time steps $t$ and $t+1$ is simply
\begin{equation}
    \delta E_{c}(t+1) - \delta E_{c}(t) =
    \begin{pmatrix}
        \delta q_a & \delta q_b
    \end{pmatrix}
    \begin{pmatrix}
        K_{ac} & K_{ad} \\
        K_{bc} & K_{bd}
    \end{pmatrix}
    \begin{pmatrix}
        \delta q_c \\
        \delta q_d
    \end{pmatrix}
    \, .
    \label{eqn:coulomb-spin-flip-local}
\end{equation}
Computing the Coulomb energy using the above expression~\eqref{eqn:coulomb-spin-flip-local} is substantially faster than~\eqref{eqn:coulomb-spin-flip} since it involves an $O(1)$ number of terms as opposed to $O(L^3)$. If one or both of the tetrahedra $a$,~$b$ and $c$,~$d$ do overlap, then the expression~\eqref{eqn:coulomb-spin-flip-local} must be modified, but it remains $O(1)$ in complexity per spin. Hence, the overall complexity scales as $O(L^3)$ per WTMC sweep.

The dipolar interaction between spins, $E_d=\sum_{i<j}S_i K_{ij} S_j$, can also be implemented in a similar way with $O(1)$ complexity per spin. Suppose that at time $t$ the spin $S_r$ was flipped, and we would like to then propose flipping $S_k$ both before and after flipping spin $S_r$. We find that in the case $k \neq r$
\begin{equation}
    \delta E_{d}^{(k)}(t+1) - \delta E_{d}^{(k)}(t) = -4 S_k(t) K_{kr} S_r(t)
    \, .
\end{equation}
In the special case $k=r$, we are proposing to reverse the previous spin flip and therefore $\delta E_{d}^{(k)}(t+1) - \delta E_{d}^{(k)}(t) = -2 \delta E_{d}^{(k)}(t)$.

The absolute values of the spin flip energies must be recomputed periodically using~\eqref{eqn:coulomb-spin-flip}, or the equivalent expression in the case of dipolar interactions between spins, in order to prevent the accumulation of numerical error.

For truncated interactions, we need not generate fresh waiting times for all the spins at each step---only those affected by the previous update~\cite{Dall2001}. Hence, the complexity in this case scales as $O(\ln L)$ per WTMC sweep, allowing much larger systems to be accessed.
%
%

\section{\label{sec:meanfield-solution}
Solution to the mean field equations
        }
Making use of the integral
\begin{equation}
  \int \mathrm{d}x \, \frac{e^{-r/(1+x)^{s}}}{(1+x)^t} =
  \frac{r^{-(t-1)/s}}{s} \Gamma\left(\frac{t-1}{s}, \frac{r}{(1+x)^{s}}\right)
  \, ,
\end{equation}
for $t>1$, we find that the full time-dependence of the noncontractible pair density may be expressed in terms of the upper incomplete gamma function $\Gamma(s, x)$ as
\begin{eqnarray}
\eta(t) &=&
  e^{- \mc{R} \Theta(t)}
	\bigg\{
	  \eta_0
		+
		\rho_q^0 \frac{\alpha\mc{K}'}{\mc{K}}
		  \left( \frac{\alpha\mc{R}}{\mc{K}_0} \right)^{-\alpha}
      e^{\alpha\mc{R}/\mc{K}_0}
\label{eqn:nc-density-solution-coulomb}
\\
&\times&
  \bigg[
	  \Gamma\bigg(
		  \alpha,
			\frac{\alpha\mc{R}}{\mc{K}_0}
			  \left( 1 + \beta \mc{K}_0 \rho_q^{0} t \right)^{(\beta-1)/\beta}
		\bigg) -
    \Gamma\bigg( \alpha,  \frac{\alpha\mc{R}}{\mc{K}_0} \bigg)
	\bigg]
\bigg\}
  \, .
\nonumber
\end{eqnarray}
Note that the behaviour of $\Theta(t)$ determines whether or not a metastable plateau appears; if $\Theta(t)$ tends to a constant for large times then the system will necessarily exhibit a plateau in the noncontractible pair density $\eta(t)$.
This function is plotted in the inset of Fig.~\ref{fig:randomwalk-coulomb-quench} for comparison with the charge model with long-range interactions.
%
%

\section{\label{app: double charges}
Double charges
        }
In this Appendix we show that the presence of double monopoles does not significantly alter the conclusions of our mean field modelling in Sec.~\ref{sec:meanfield} of the main text.
In particular, we show by explicitly solving the mean field equations governing the density of monopoles subject to truncated interactions in the presence of double charges that, although the value of the plateau (in a finite system) is altered, the finite size scaling exponent $\nu$ remains unchanged.
We argue that this feature is true more generally---further modifications of the mean field equations may change the short-time dynamics of the free monopole density, but leave its asymptotic decay ($\propto 1/t$) unchanged.
This implies that the exponents derived in Sec.~\ref{sec:meanfield} are in some sense universal, while the precise value of the plateau is not (by universal we mean that the exponents
are independent of how precisely the system is prepared, and are robust to the addition of
terms in the mean field equations that lead to modifications of the
short-time dynamics).

\begin{figure}
    \centering
    \includegraphics[width=\linewidth]{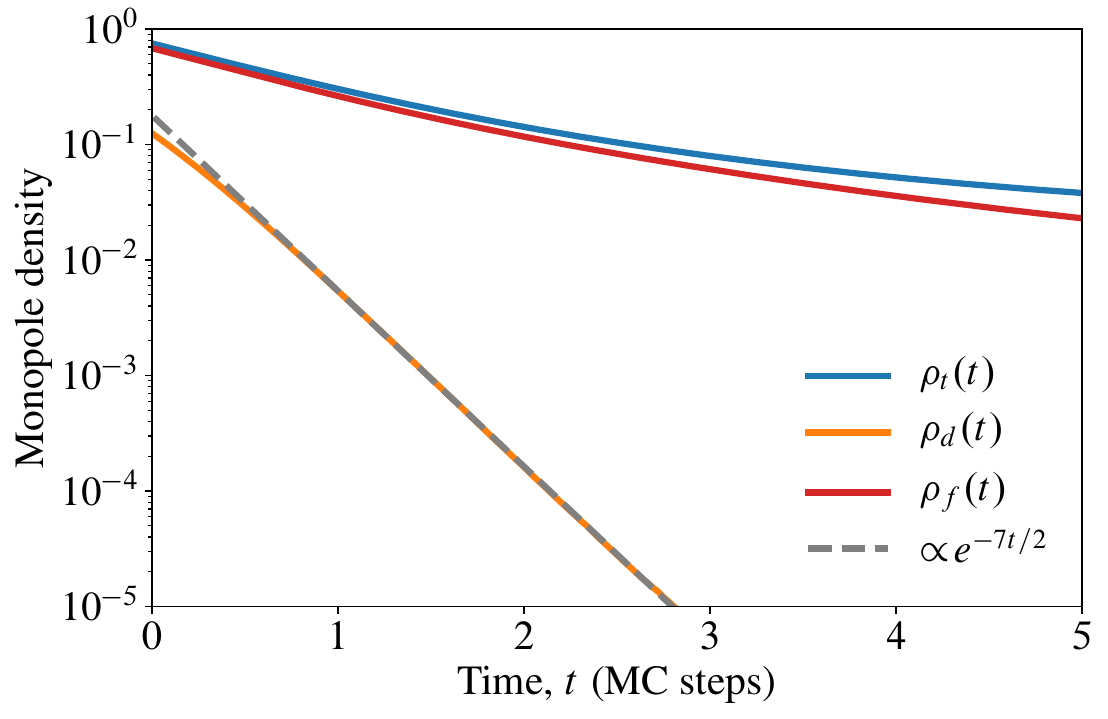}
    \caption{Decay of the various monopole densities for a thermal quench from infinite temperature down to $T = \SI{0.06}{\kelvin}$ in spin ice (system size $L=20$, i.e., $\num{128000}$ spins).
    The double charge density $\rho_d(t)$ decays exponentially with time with rate constant $\mc{K}_d = 7/2$. At very short times, $t \lesssim 1$, the effect of a nonzero free monopole density cannot be neglected, and the rate of double charge decay is reduced due to obstructed decay channels.}
    \label{fig:double-charge-exponential-decay}
\end{figure}

In addition to the species considered in Sec.~\ref{sec:meanfield}, we introduce two new densities, $d_q(t)$ (where $q=\pm$), which equal the fraction of sites that host a charge $Q=\pm 2$, respectively.
Notice that a double charge can always decay by reacting with any of its neighbouring tetrahedra (be them empty, occupied by a single or by a double charge), with the only exception being when it neighbours a single charge of the same sign, in which case flipping the intervening spin merely swaps the single and double charge without annihilating either of them.
In principle the time evolution of the double charges depends therefore on the evolution of the single monopole density.
Indeed, the average number of bonds surrounding an isolated double charge $2q$ along which it is able to decay is $4(1-\rho_q)$ at the mean field level, i.e., assuming that each site is independent.
The \emph{asymptotic} decay of the double monopole density is however determined by neighbouring double charges of opposite sign since the number of bonds along which the pair may decay is $7/2 - 3(\rho_q + \rho_{\bar{q}})/2$ per site.
Therefore, for all but the shortest times where the effect of nonzero $\rho_q$ cannot be neglected, we expect the double charge density to decouple from the other monopole densities and to decay exponentially with a rate constant $\mc{K}_d \simeq 7/2$, i.e.,
\begin{equation}
  \dv{d_q}{t} = -\mc{K}_d d_q
  \, .
  \label{eqn:double-charge-decay}
\end{equation}
Adding the two equations for $q=\pm$, we obtain $\rho_d(t) = \rho_d^0 e^{-\mc{K}_d t}$.
This expectation is indeed confirmed by our Monte Carlo simulations of CSI, where we observe asymptotic exponential decay of the total double charge density $\rho_d(t) = d_+(t) + d_-(t)$ with time (see Fig.~\ref{fig:double-charge-exponential-decay}), consistent with the prediction $\mc{K}_d = 7/2$.

The equation governing the free charge density $\rho_q$ must also be modified to include the effect of double monopole decay:
\begin{equation}
  \dv{\rho_q}{t} = -\mc{K}\rho_+\rho_- +2\mc{K}_d^\prime d_q(t)
  \, .
  \label{eqn:free-charge-decay}
\end{equation}
The rate constant $\mc{K}_d'$ corresponds to the spontaneous decay channel into adjacent empty sites only, implying that $\mc{K}_d' < \mc{K}_d$.
Hence, the effect of including a nonzero density of double charges on the free monopole density is to add an exponentially decaying source term that corresponds to the production of free monopoles when double charges decay spontaneously.
If we took into account spatial fluctuations, then we would also need to include a term $\propto(d_q\rho_{\bar{q}} - d_{\bar{q}}\rho_q)$ in this equation, but at the mean field level, charge neutrality of the single and double charges \emph{separately} implies perfect cancellation of such a term. That is, when a single free charge $q$ meets a double charge $2\bar{q}$, a free charge $q$ is removed and a free charge $\bar{q}$ is created. However, the rate at which this process occurs is identical for $q=\pm$.
Substituting the exponential decay of $\rho_d(t)$ into this equation, we must solve the nonlinear equation
\begin{equation}
  \dv{\rho_q}{t} + \mc{K}\rho_q^2 = \mc{K}_d^\prime \rho_d^0 e^{-\mc{K}_d t}
  \, ,
\end{equation}
for $\rho_q(t)$, in which we have made use of charge neutrality, $\rho_+(t) = \rho_-(t)$.
This equation has the exact solution
\begin{equation}
  \rho_q(t) = y \frac{\mc{K}_d}{2\mc{K}}  \frac{K_1(y)-cI_1(y)}{K_0(y)+cI_0(y)}
  \, ,
\end{equation}
where we have written, for convenience of notation, $y(t) \equiv 2 \sqrt{\mc{K}\mc{K}_d^\prime\rho_d^0/\mc{K}_d^2} e^{-\mc{K}_d t / 2}$. The constant $c$ is determined by the initial conditions $\rho_q(0)=\rho_q^0$, and $I_n(x)$ and $K_n(x)$ are modified Bessel functions of the first and second kind, respectively.

Finally, the expression for $\eta(t)$ must also be modified for direct comparison with our numerical results. When two double charges (of opposite sign) are adjacent to one another, the bond necessarily hosts one contractible pair and one noncontractible pair.
The number of adjacent doubly occupied sites is simply proportional to $\rho_d(t)$ at long times, and the corresponding contribution to $\eta(t)$ contributes towards the kink in the noncontractible pair density observed in our numerical simulations at the characteristic decay time $t\sim\mc{K}_d^{-1}$ of the double charges.
At later times, the equation for $\eta(t)$ remains unchanged~\footnote{Including terms that correspond to the decay of neighbouring double charges into free monopoles gives rise to an exponentially decaying contribution to $\eta(t)$.}
\begin{equation}
  \dv{\eta}{t} = -\frac12 \mc{R}(\rho_+ + \rho_-) \eta + \mc{K}'\rho_+\rho_-
  \, .
\end{equation}
The form of the solution is
\begin{equation}
  \eta(t) = e^{-\mc{R}\Theta(t) } \left[ \eta(0) - \frac{\mc{K}'}{\mc{K}} \int_0^{t} \mathrm{d}t^\prime e^{\mc{R}\Theta(t')}  \dot{\rho}_q(t')   \right]
  \, ,
  \label{eqn:eta1-solution}
\end{equation}
where we recall that $\Theta(t) \equiv \int_0^t \mathrm{d}t'\, \rho_q(t')$.
Hence, the asymptotic behaviour of $\eta(t)$ is directly determined by the asymptotic behaviour of $\rho_q(t)$.
In order to derive this behaviour, we require the expansions of $I_n(x)$ and $K_n(x)$ for small values of the argument $x$~\cite{Abramowitz1965}:
\begin{align}
  I_0(x) &= 1 + O(x^2) \, , \\
  I_1(x) &= \frac{1}{2}x + O(x^3) \, , \\
  K_0(x) &= -\ln \frac{e^\gamma}{2} x + O(x^2\ln x) \, , \\
  K_1(x) &= \frac{1}{x} + \frac{1}{2}x\ln x + O(x) \, ,
\end{align}
where $\gamma\simeq 0.5772$ is the Euler--Mascheroni constant. These expansions allow us to deduce that
\begin{align}
  \rho_q(t) &= \frac{\mc{K}_d}{2\mc{K}}y \frac{1/y + (y/2) \ln y + O(y)}{ \ln 2 -\ln e^\gamma y  + c + O(y^2 \ln y)} \\
            &= \frac{1}{\mc{K}t} + O\left( t^{-2} \right)
  \, , \label{eqn:asymptotic-expansion-free-density}
\end{align}
independent of the initial conditions and independent of the initial rapid decay of double monopoles.
The subleading term $\propto 1/t^2$ depends on the short-time dynamics through $\log y_0$ and through $c$.
Correspondingly, for sufficiently large times,
\begin{equation}
  \eta(t) \propto
  \frac{1}{(\mc{K}\rho_q^0 t)^{\mc{R}/\mc{K}}}
  \, .
  \label{eqn:renormalised-eta-decay}
\end{equation}
The constant of proportionality is slightly renormalised in the presence of double charges since the asymptotic expansion of the second term in~\eqref{eqn:eta1-solution} depends on $\int_0^\infty \mathrm{d}t\, e^{\mc{R}\Theta}\dot{\rho}_q$, which in turn depends on the full time-dependence of $\rho_q(t)$, including its short-time dynamics.
However, the \emph{exponent} $\nu$ is insensitive to such details [being determined by the exponents of the leading terms in~\eqref{eqn:asymptotic-expansion-free-density} and~\eqref{eqn:renormalised-eta-decay}], and the scaling arguments presented in the main text remain robust to the addition of doubly occupied sites.
That is, the precise value of the plateau is sensitive to the addition of double monopoles into the model, but the finite size scaling exponent $\nu = 3\mc{R}/\mc{K}$ remains unchanged.

Similarly, when the charges are subject to mutual Coulombic interactions, if the leading term in the asymptotic expansion of $\rho_q(t)$ remains proportional to $t^{-\beta^{-1}}$, then the leading, time-independent term in $\Theta(t) = \text{const.} + O(t^{-(1-\beta)/\beta})$ \emph{will} be sensitive to the presence of double charges.
Therefore, since this term contributes to the value of the plateau in the thermodynamic limit, $\eta_\infty$ from~\eqref{eqn:thermodynamic-plateau} will be modified slightly in the presence of doubly occupied sites.
However, the subleading contribution ($\sim t^{-(1-\beta)/\beta}$), which determines the finite size scaling exponent $\nu$, will again be robust to the addition of doubly occupied sites, and the relation $\nu=3(1-\beta)$, which relates the asymptotic decay of $\rho_q$ to the finite size scaling behaviour, also remains unchanged.

More generally, adding further terms to our mean field equations (which depend on higher powers of the various densities) will indeed modify the short-time dynamics of $\rho_q(t)$. The precise density at which the plateau occurs in a system of finite size in the case of truncated interactions, and the value of the plateau in the thermodynamic limit in the case of long-range interactions depend---through~\eqref{eqn:eta1-solution}---on the \emph{full} history of $\rho_q(t)$, and therefore will be modified. However, the \emph{asymptotic} behaviour of $\rho_q(t)$, which directly determines the finite size scaling exponent $\nu$ for both types of interaction, is insensitive to such details.
%
%

\section{\label{app: blocked directions}
Blocked directions
        }
To derive the probability that a given monopole is pinned, it is convenient to use the following convention for the normalised basis vectors:
\begin{align}
  \v{e}_0 &= \v{e}_z \, , \\
  \v{e}_1 &= \tfrac{2\sqrt{2}}{3} \v{e}_x - \tfrac{1}{3}\v{e}_z \, , \\
  \v{e}_2 &= -\tfrac{\sqrt{2}}{3}( \v{e}_x + \sqrt{3} \v{e}_y ) - \tfrac{1}{3}\v{e}_z \, , \\
  \v{e}_3 &= -\tfrac{\sqrt{2}}{3}( \v{e}_x - \sqrt{3} \v{e}_y ) - \tfrac{1}{3}\v{e}_z
  \, .
\end{align}
Now, the probability that a given monopole is instantaneously pinned, $p_b$, is $\Omega_b/4\pi$, where $\Omega_b$ is the solid angle over which there exists a positive projection onto exactly one of $\v{e}_\mu$ ($\mu=0$--$3$).
In this case, there exists only one direction which lowers the energy of the monopole, and so the monopole will be pinned if the minority spin coincides with this direction.

For convenience, let us consider the solid angle $\Omega_0$ corresponding to a positive projection onto $\v{e}_0$, and a negative projection onto the remaining three basis vectors. By symmetry, $\Omega_b = \Omega_0$. We therefore require that the following conditions are simultaneously satisfied
\begin{align}
  \cos \theta &> 0 \, , \label{eqn:solid-angle-1} \\
  2\sqrt{2} \sin\theta\cos\phi - \cos\theta &< 0 \, , \label{eqn:solid-angle-2} \\
  \sqrt{2} (-\sin\theta \cos\phi - \sqrt{3}\sin\theta\sin\phi) - \cos\theta &< 0 \, , \label{eqn:solid-angle-3} \\
  \sqrt{2} (-\sin\theta \cos\phi + \sqrt{3}\sin\theta\sin\phi) - \cos\theta &< 0 \, , \label{eqn:solid-angle-4}
\end{align}
where we have parameterised the unit sphere using polar and azimuthal angles $\theta$ and $\phi$, respectively. The corresponding solid angle defined by this region is (taking advantage of the $D_3$ symmetry about the $z$-axis)
\begin{align}
  \Omega_0 &= 6\int_0^{\pi/3}\mathrm{d}\phi\, \int_0^{f(\phi)} \mathrm{d}\theta \, \sin\theta \\
  &= 6\int_0^{\pi/3} \mathrm{d}\phi\, \left[ 1 - \cos f(\phi) \right] \, ,
  \label{eqn:solid-angle-single}
\end{align}
where $f(\phi)$ is defined implicitly by the condition $2\sqrt{2}\sin f(\phi)\cos\phi - \cos f(\phi)=0$, i.e., the limiting case of condition~\eqref{eqn:solid-angle-2}. The other conditions~\eqref{eqn:solid-angle-3} and~\eqref{eqn:solid-angle-4} are also automatically satisfied if~\eqref{eqn:solid-angle-2} is satisfied in the region $0 < \phi < \pi/3$. Hence,
\begin{equation}
  \cos f(\phi) = \frac{2\sqrt{2}\cos\phi}{\sqrt{1 + (2\sqrt{2}\cos\phi)^2}}
  \, ,
\end{equation}
and the integral~\eqref{eqn:solid-angle-single} over the azimuthal angle $\phi$ may be evaluated exactly to give
\begin{equation}
  \Omega_0 = 6\left[ \frac{\pi}{3} - \arctan\sqrt{2} \right]
  \, ,
\end{equation}
and finally $p_b = \Omega_0/4\pi$, giving the result stated in the main text.
%
%

\section{\label{app:finite-size-exponents}
Corrections to the finite size scaling exponent for CSI with truncated interactions
        }

\begin{figure}[t]
\subfloat[$f=6/7$\label{fig:blocking_probabilities:a}]{\includegraphics[width=0.33\linewidth,valign=c]{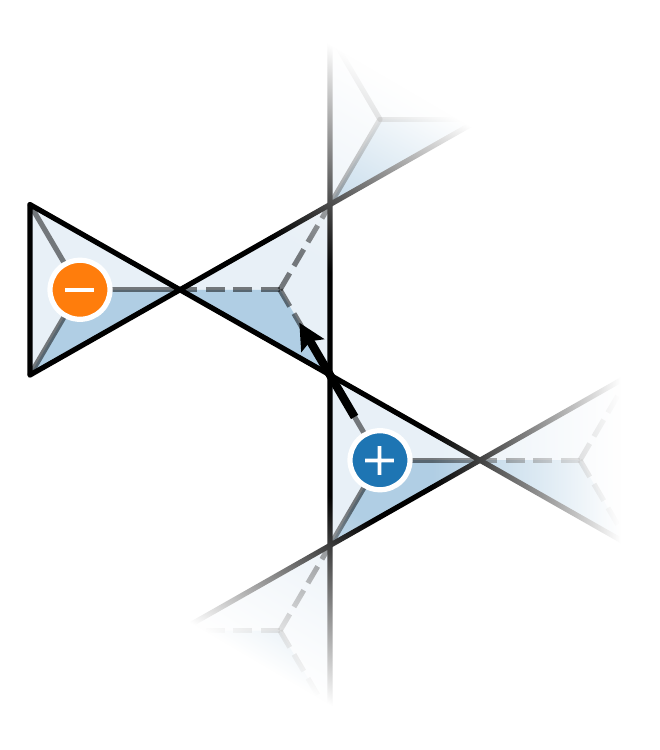}}%
  \hfill
\subfloat[$f=3/5$\label{fig:blocking_probabilities:b}]{\includegraphics[width=0.33\linewidth,valign=c]{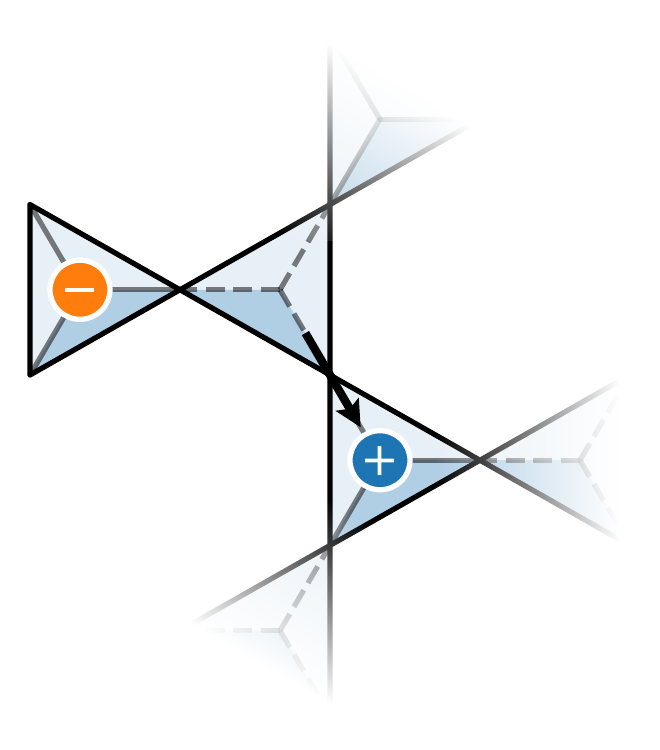}}%
  \hfill
\subfloat[$f=0$\label{fig:blocking_probabilities:c}]{\includegraphics[width=0.33\linewidth,valign=c]{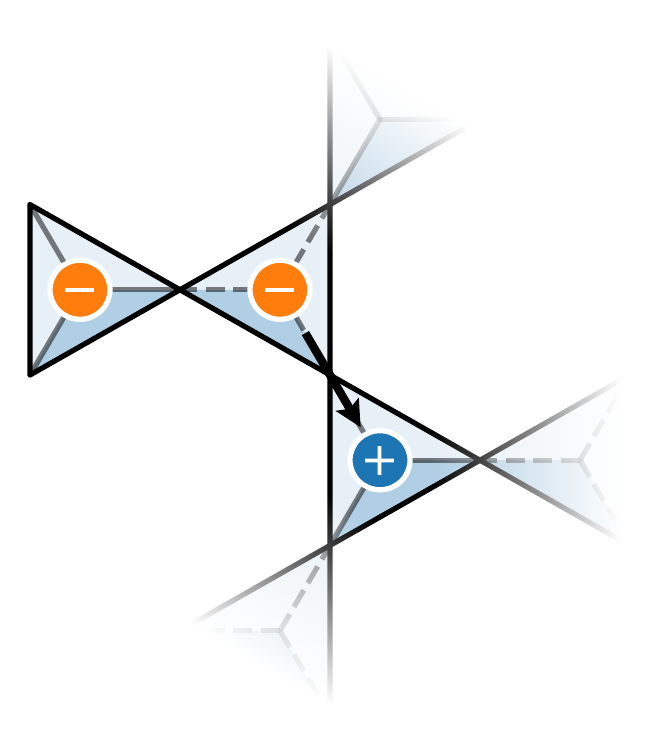}}%
  \caption{Fraction $f$ of spin configurations in which the positively and
  negatively charged monopoles are able to annihilate (or form a new
  noncontractible pair), for different configurations of spins on the central
  tetrahedron (which hosts the positively charged monopole).}
  \label{fig:blocking_probabilities}
\end{figure}

Here we show how the finite size scaling exponent $\nu$, which determines the
finite size scaling behaviour of the plateau in classical spin ice with truncated interactions between the monopoles (Section~\ref{sec:meanfield:truncated}), $\eta_p(L) \sim L^{-\nu}$, is
affected by the inclusion of blocked directions.

As shown in Sec.~\ref{sec:summary}, the expression for the exponent $\nu$ is given in terms of the
ratio of the rate of monopole-assisted decay, $\mc{R}$, to the rate of monopole--antimonopole
collision events, $\mc{K}$ (during which
the two monopoles either annihilate or form a new noncontractible pair); specifically, $\nu = 3\mc{R}/\mc{K}$.
In order to estimate the ratio $\mc{R}/\mc{K}$ microscopically, we consider a
symmetrical cluster consisting of a central tetrahedron, and its first and
second neighbouring tetrahedra (considering only the first nearest neighbours reproduces
$\mc{R}/\mc{K}=3/4$, i.e., blocked directions have no effect at this level).
For concreteness, suppose that the central tetrahedron hosts a single,
positively charged monopole.
This positively charged monopole is either (1) isolated, corresponding to the
calculation of the rate constant $\mc{K}$, or (2) one half of a noncontractible
pair, with its negatively charged partner sitting on one of the first
nearest-neighbour tetrahedra, corresponding to the calculation of $\mc{R}$.
Assuming that there is an equal probability of finding the negatively
charged monopole on any of the second-neighbour sites, we then deduce that
the ratio of rates equals
\begin{equation}
  \frac{\mc{R}}{\mc{K}} = \frac{{P}_2}{{P}_1}
  \, ,
\end{equation}
where $P_{n}$ ($n=1,2$) is the probability that a negatively charged
monopole situated at second-neighbour distance
is able to reach the positively charged monopole at the central site and
annihilate (for $n=2$) or either annihilate or form a new noncontractible pair
(for $n=1$).

The probabilities $P_n$ may then be computed by enumerating the possible
spin configurations:
\begin{align}
  P_1 &= \frac{1}{N} \sum_{i=1}^N f_i = \frac14 \left[ 3\times \frac67 + \frac35 \right] = \frac{111}{140} \\
  P_2 &= \frac{1}{N} \sum_{i=1}^N f_i  = \frac14 \left[ 3\times \frac67 + 0 \right] = \frac{9}{14}
  \, ,
\end{align}
where $N \equiv 12$ is the number of second neighbours, and $f_i$
is the fraction of paths starting on site $i$ that can reach the
central tetrahedron.
In both expressions, the factor $6/7$ corresponds to the fraction of paths
that are not blocked when the negatively charged monopole is approaching a
vertex on the central tetrahedron that hosts a majority spin (Fig.~\ref{fig:blocking_probabilities:a}).
Conversely, $3/5$ of the paths are not blocked when approaching the
minority spin (Fig.~\ref{fig:blocking_probabilities:b}), unless the first-neighbour site is occupied by
the negatively charged member of a noncontractible pair (Fig.~\ref{fig:blocking_probabilities:c}), in which
case the fraction is zero.

Using these probabilities, we obtain the ratio of rates
\begin{equation}
  \frac{\mc{R}}{\mc{K}} = \frac{30}{37}
  \, .
\end{equation}
Finally, the finite size scaling exponent is therefore $\nu = 90/37$.

%
%

\section{\label{app:J2-J3}
Farther-ranged interactions
        }

\begin{figure}[t]
    \centering
    \includegraphics[width=0.65\linewidth]{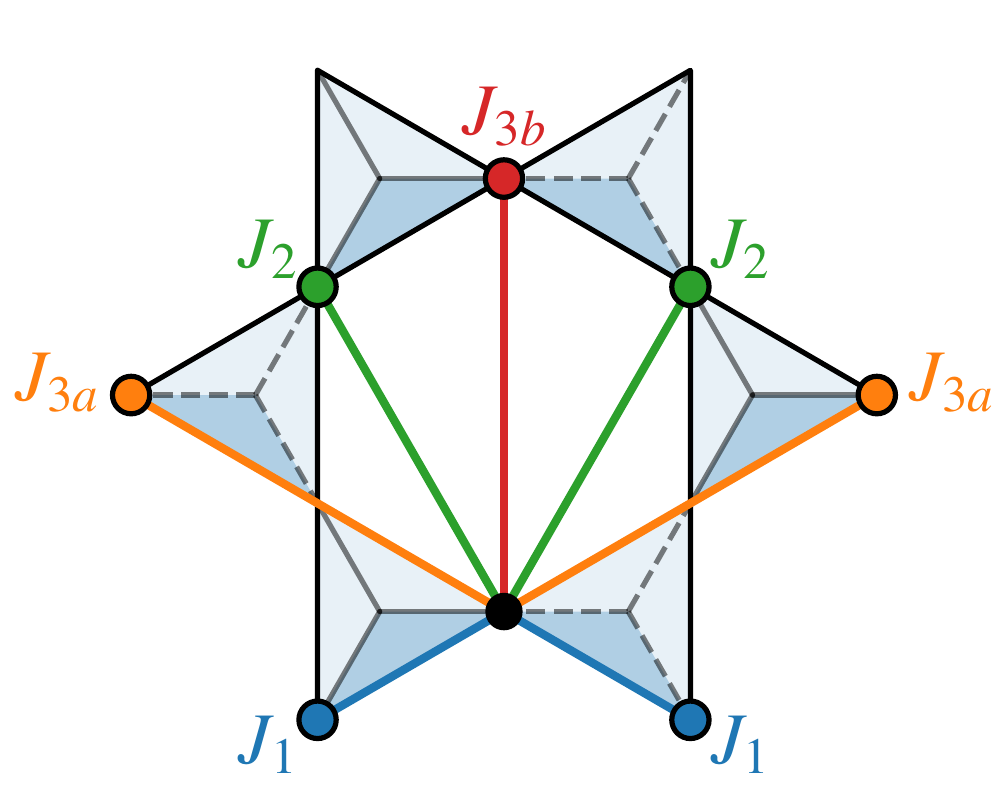}
    \caption{Illustration of the various
    coupling constants $J_1$ (blue), $J_2$ (green), $J_{3a}$ (orange)
    and $J_{3b}$ (red). The solid lines
    denote interactions between the central spin, represented by the
    solid black circle, and its neighbouring spins, represented by the coloured
    circles.}
    \label{fig:spin-spin-interactions}
\end{figure}

Consider the generalised dipolar spin ice Hamiltonian, which includes
interactions $J_{ij}$ between spins beyond nearest-neighbour separation~\cite{Yavorskii2008, Henelius2016, Borzi2016, Samarakoon2019}:
\begin{equation}
  H = -\sum_{(ij)} J_{ij}  \v{S}_i \cdot \v{S}_j + D \sum_{(ij)} \left[ \frac{\v{S}_i\!\cdot\!\v{S}_j}{|\v{r}_{ij}|^3} - \frac{3(\v{S}_i\!\cdot\!\v{r}_{ij})(\v{S}_j\!\cdot\!\v{r}_{ij})}{|\v{r}_{ij}|^5} \right]
  \, , 
\end{equation}
where $\v{S}_i = S_i \v{e}_i$, with $S_i \in \{ -1, +1 \}$.
Let us focus in particular on $J_2$- and $J_3$-type interactions, of
the form shown in Fig.~\ref{fig:spin-spin-interactions}, where the latter
is divided into $J_{3a}$ and $J_{3b}$, corresponding to spin--spin interactions between adjacent hexagons and across hexagonal plaquettes, respectively. These interactions may be written in terms of the Ising spins $S_i$ as follows:
\begin{align}
    -\sum_{(ij)} J_{ij} \v{S}_i \cdot \v{S}_j &= \frac{J}{3} \sum_{\langle ij \rangle_1}   S_i S_j + \frac{J_2}{3} \sum_{\langle ij \rangle_2} S_i S_j \nonumber \\
&- J_{3a} \sum_{\langle ij \rangle_{3a}} S_i S_j - J_{3b} \sum_{\langle ij \rangle_{3b}} S_i S_j
\, ,
   \label{eqn:farther-ranged-exchange-general}
\end{align}
where $\langle ij \rangle_n$ denotes $n$th neighbours on the pyrochlore lattice. In the special case $J_2/3 + J_{3a} = J_{3b} = 0$, the spin--spin interactions assume the form
\begin{multline}
    -\sum_{(ij)} J_{ij} \v{S}_i \cdot \v{S}_j = \\
    \frac23 \left( J -2 J_2 \right)\sum_{a} Q_a^2+ \frac{4J_2}{3} \sum_{\langle ab \rangle} Q_a Q_b + \text{const.}
    \label{eqn:farther-ranged-exchange}
\end{multline}
That is, the farther-ranged interactions between spins may be summed
to give nearest-neighbour truncated interactions between the tetrahedral
charges $Q_a$, in addition to a shift in the chemical potential for
monopoles (i.e., a shift in the value of $2J_\text{eff}$). As a result,
only the short-distance physics of monopole dynamics is affected, leading to
modifications of, for example, the energy barrier associated with thermally activated decay
of noncontractible pairs.

When $J_{2} / 3 + J_{3a} \neq 0$ or $J_{3b} \neq 0$, the spin--spin interactions in \eqref{eqn:farther-ranged-exchange-general}
can no longer be written in terms of tetrahedral charges $Q_a$ only,
leading to a form of effective disorder in the dynamics of monopoles.
However,
as long as the long-range bias for monopole motion across the system is
active, we expect the phenomenology discussed in the main text to remain
qualitatively similar. 
For example, if the magnitude of the interactions $J_2$ and $J_3$ is significantly
smaller than $E_\text{nn}$, the dynamics of
the monopoles remains essentially unaffected during the transient regime
in which the plateau is established. 
This is indeed the case in {\DTO}, where estimates of $J_2$ and
$J_3$ typically range from $O(\SI{1}{\milli\kelvin})$ to $O(\SI{10}{\milli\kelvin})$~\cite{Yavorskii2008, Henelius2016, Borzi2016, Samarakoon2019}.

%
%

\bibliographystyle{aipnum4-1}
\bibliography{references}


\end{document}